\documentclass{article}

\usepackage{arxiv}

\usepackage[utf8]{inputenc} 
\usepackage[T1]{fontenc}    
\usepackage[english]{babel}
\usepackage{microtype}      

\usepackage{tgpagella}      

\usepackage{amsfonts}       
\usepackage{amssymb}
\usepackage{amsmath}
\usepackage{bm}             
\usepackage{nicefrac}       

\usepackage{enumitem}

\usepackage[ruled,vlined]{algorithm2e}

\usepackage[numbers]{natbib}
\usepackage{doi}

\usepackage[dvipsnames,table]{xcolor}
\usepackage{graphicx}
\usepackage[caption=false]{subfig}  
\usepackage{booktabs}       
\usepackage{multirow}       
\usepackage{longtable}      
\usepackage{array}          

\usepackage{hyperref}       
\usepackage{url}            

\title{Unconventional application of k-means for distributed approximate similarity search}

\author{ \href{https://orcid.org/0000-0003-0231-2051}{\includegraphics[scale=0.06]{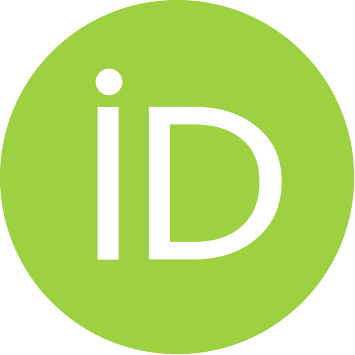}\hspace{1mm}Felipe Ortega}\thanks{Corresponding author. This article is currently under peer-review. All datasets are either publicly available or can be completely reproduced following the guidance provided in the associated references.} \\
	Data Science Lab (DSLAB), CETINIA\\
	Rey Juan Carlos University\\
	Camino del Molino, 5. 28943\\
	Fuenlabrada, Madrid. SPAIN\\
	\texttt{felipe.ortega@urjc.es} \\
	\And
	\href{https://orcid.org/0000-0002-7539-8522}{\includegraphics[scale=0.06]{orcid.pdf}\hspace{1mm}Maria Jesus Algar} \\
	Data Science Lab (DSLAB), CETINIA\\
	Rey Juan Carlos University\\
	Tulipán s/n. 28933\\
	Móstoles, Madrid. SPAIN\\
	\texttt{mariajesus.algar@urjc.es} \\
	\And
	\href{https://orcid.org/0000-0001-5197-2932}{\includegraphics[scale=0.06]{orcid.pdf}\hspace{1mm}Isaac Martín de Diego} \\
	Data Science Lab (DSLAB), CETINIA\\
	Rey Juan Carlos University\\
	Tulipán s/n. 28933\\
	Móstoles, Madrid. SPAIN\\
	\texttt{isaac.martin@urjc.es} \\
	\And
	\href{https://orcid.org/0000-0003-1415-1961}{\includegraphics[scale=0.06]{orcid.pdf}\hspace{1mm}Javier M. Moguerza} \\
	Data Science Lab (DSLAB), CETINIA\\
	Rey Juan Carlos University\\
	Tulipán s/n. 28933\\
	Móstoles, Madrid. SPAIN\\
	\texttt{javier.moguerza@urjc.es} \\
}

\renewcommand{\headeright}{Preprint article}
\renewcommand{\undertitle}{Preprint article}
\renewcommand{\shorttitle}{Unconv. app. of k-means for distrib. approx. similarity search}

\hypersetup{
    pdftitle={A template for the arxiv style},
    pdfsubject={q-bio.NC, q-bio.QM},
    pdfauthor={David S.~Hippocampus, Elias D.~Striatum},
    pdfkeywords={First keyword, Second keyword, More},
    colorlinks=true,        
    linkcolor=Sepia,      
    citecolor=ForestGreen,    
    urlcolor=NavyBlue,      
    linktocpage=true        
}

\begin{document}
\maketitle

\begin{abstract}
    Similarity search based on a distance function in metric spaces is a 
    fundamental problem for many applications. Queries for similar objects lead 
    to the well-known machine learning task of nearest-neighbours identification. 
    Many data indexing strategies, collectively known as Metric Access 
    Methods (MAM), have been proposed to speed up queries for similar elements
    in this context. Moreover, since exact approaches to solve similarity queries 
    can be complex and time-consuming, alternative options have appeared to reduce 
    query execution time, such as returning approximate results or resorting to 
    distributed computing platforms. In this paper, we introduce MASK (Multilevel
    Approximate Similarity search with $k$-means), an unconventional application of the
    $k$-means algorithm as the foundation of a multilevel index structure for 
    approximate similarity search, suitable for metric spaces. We show that inherent 
    properties of $k$-means, like representing high-density data areas with fewer 
    prototypes, can be leveraged for this purpose. An implementation of this new 
    indexing method is evaluated, using a synthetic dataset and a real-world dataset 
    in a high-dimensional and high-sparsity space. Results are promising and 
    underpin the applicability of this novel indexing method in multiple domains.
\end{abstract}



\section{Introduction}
\label{intro}

\textit{Similarity search}, also known as \textit{proximity search},
\citep{Chavez2001,Samet2006,Zezula2006} is a cornerstone for 
applications in many different fields such as databases 
\citep{Petrov2019dbs}, information retrieval \citep{baeza2011modern}, 
distributed data processing \citep{Cech2020}, computer vision~\citep{Muja2014} and bioinformatics~\citep{Tuncbag2011,Liu2020covid}, 
among others. Elements from a dataset are represented by 
\textit{feature vectors} in a multidimensional space, and
the goal is to find which elements are similar (close) to a 
given query object, subject to a certain measure of similarity or,
conversely, dissimilarity.

More formally, let $\mathcal{X}$ be the domain of elements
represented by their descriptive features and $s: \mathcal{X} \times 
\mathcal{X} \mapsto \mathbb{R}$ a similarity function that, for each pair
of elements $a, b \in \mathcal{X}$, returns a real number $s(a,b)$ representing
a similitude score between these two elements. In some cases, it is more
convenient to define an equivalent dissimilarity function $\delta: \mathcal{X} \times 
\mathcal{X} \mapsto \mathbb{R}$, so that for $a, b, c \in \mathcal{X}$ it
holds that $s(a,b) > s(a,c) \Leftrightarrow \delta(a,b) < \delta(a,c)$. The
pair $(\mathcal{X}, \delta)$ is known as a \textit{dissimilarity space}, a kind
of topological space~\citep{Duin2012}. In addition, if $\delta$ satisfies the
properties $\delta(a,b) \geq 0$ (non-negativity), $\delta(a,a) = 0$ (reflexivity),
$\delta(a,b) = \delta(b,a)$  (symmetry) and
$\delta(a,b) + \delta(b,c) \geq \delta(a,c)$ (triangle inequality), then 
the dissimilarity function is a \textit{metric} and it is usually termed as a
\textit{distance} \citep{Duda2001,webb2011}.
\smallskip

Given a query object $q \in \mathcal{X}$, the goal of similarity 
search is to resolve some of the following queries~\citep{Samet2006}:

\begin{itemize}
    \item \textit{Point query}: Finding elements in $\mathcal{X}$ with
    exactly the same feature values as $q$.
    \smallskip
    
    \item \textit{Range query}: Retrieving a subset of elements 
    $\{o_i\} \in \mathcal{X}$
    whose feature values lie within the scope of a given similarity
    threshold $r$ around $q$, so that $\forall\,o_i,\;\delta(o_i, q) \leq r$.
    \smallskip
    
    \item \textit{Nearest-neighbour query}: Recovering elements in
    $\mathcal{X}$ whose features are the most similar to the feature
    vector representing $q$. In this case, we may be interested in
    finding the single most similar element to $q$ (\textit{nearest
    neighbour}), denoted by $NN_q$, or the $k$ closest elements to $q$
    (\textit{k-nearest neighbours}), denoted by $k{\text -}NN_q$.
    
\end{itemize}

Numerous data indexing methods, collectively known as \textit{access methods}, 
have been proposed to speed up similarity queries. These methods create an 
index structure to partition the whole domain $\mathcal{X}$ in different 
regions, according to a distance function. After this, users employ different 
search algorithms to solve similarity queries using this index. There is a
wide range of access methods, including exact approaches, approximate solutions
and variants tailored to distributed computing systems.
\smallskip

Exact similarity search methods identify the completely accurate result 
set for a query. In vector spaces, multi-attribute access 
methods use the absolute position of objects to group elements and 
search for similar instances. In a \textit{d}-dimensional 
Euclidean space, these indexing structures are jointly known 
as \textit{multidimensional access methods}. These can be classified
as Point Access Methods (PAM), for elements without a spatial
extension, and Spatial Access Methods (SAM), to search extended 
elements such as lines, polygons or higher-dimensional polyhedra.
In research literature, the general acronym SAM commonly 
designates both classes of indexing strategies in vector spaces~\citep{baeza2011modern}.
\medskip

However, SAM present several limitations. First, in applications like
textual and multimedia databases, bioinformatics or pattern recognition,
elements in $\mathcal{X}$ cannot always be described in a vector space.
Second, to compare the similarity between any two elements
we must use a distance function that avoids introducing any correlation between
feature values~\citep{Faloutsos1994}. A general distance function that meets
this requirement is the Minkowski distance~\citep{Flach2012}.
The \textit{Manhattan} or \textit{city block} distance (\(L_1\) norm), 
the \textit{Euclidean} distance (\(L_2\) norm) and the \textit{Chebyshev}
distance (\(L_{\infty}\) norm) are typical instances of this family.
Third, SAM are prepared for data with spatial components or represented in
a vector space with a relatively low number of dimensions. In high-dimensional 
spaces their data partition algorithms become unusable~\citep{Bohm2001},
due to the \textit{curse of dimensionality} \citep{Bellman2015}.
Hence, all elements tend to be very far from each other, disregarding the 
distance function chosen for the index.  Feature
selection \citep{Kuhn2020FeatEng} or dimensionality reduction techniques
like multidimensional scaling~\citep{Hastie2009} can mitigate this
problem. However, these approaches will only be useful if the effective number of 
dimensions in which the elements are actually represented is low. The so-called
\textit{intrinsic dimensionality} of a dataset \(\mathcal{X}\), accounts 
for this notion of ``effective dimensions'', and is given by
\(\rho = \frac{\mu^2}{2\sigma^2}\), where \(\mu\) and \(\sigma^2\) 
are the mean and variance of the distribution of distance values 
between any pair of elements in \(\mathcal{X}\), respectively.
\medskip

Metric Access Methods (MAM) provide a more general framework for
similarity search problems set out in metric spaces \citep{Chavez2001,Zezula2006}.
If $d: \mathcal{X} \times \mathcal{X} \mapsto \mathbb{R}$ is a distance 
function (hence, a metric), then the pair \((\mathcal{X},d)\) defines 
a metric space. In this setting, MAM exploit the triangle inequality to 
partition the metric space into subspaces, so that they can filter out
portions of the dataset that cannot contain valid results.
This is a more general framework that subsumes SAM, since every 
normed vector space induces a metric. Additionally, alternative frameworks 
can leverage different properties, like Ptolemaic Access Methods (PtoAM)
\citep{Hetland2013}, that substitute the triangle inequality for 
Ptolemy's inequality and use distances where it is applicable or
Supermetric Search~\citep{Connor2019}, which involves
semimetric spaces where the so-called four-point property holds.

Nevertheless, MAM also present limitations. In certain complex problems 
there is no topological information about the application domain, so that
only non-metric similarity or dissimilarity functions are
available \citep{Skopal2011}. Examples include cosine 
similarity in information retrieval, dynamic time warping in time 
series and several edit distances in computational biology. This
lack of information about the problem context impedes the development
of MAM for those cases. In addition, exact search provided by MAM
can be expensive, regarding computation time spent in the search
process and updating the index structure in presence of dynamic data.
Moreover, in high-dimensional problems MAM indexes have serious issues
to narrow down the search of candidate elements for a query. As
a result, these methods usually default to a sequential scan. 
In this situation users will accept a trade-off solution, sacrificing 
accuracy for a significant reduction in search time.
\smallskip

Approximate indexing methods implement compromise solutions for similarity
search that may introduce some error in results, although not necessarily. 
In exchange, they provide faster response time by computing fewer distance 
values or decreasing usage of computing resources to complete the search. 
These methods are aimed at either vector spaces or metric spaces
\citep{Samet2006,Zezula2006,Patella2009appsearch}.
In this paper, we introduce MASK (Multilevel Approximate Similarity search
with $k$-means), a novel indexing method based on a multilevel design 
for approximate similarity search. This method involves an unconventional 
application of the $k$-means partitioning algorithm \citep{MacQueen1967,Lloyd1982}, 
that quickly reduces the number of regions to be checked when searching 
for results. For each level in the index structure, instead of fixing a low 
value for $k$ (number of centroids), as it would be customary in clustering
problems, we force $k$-means to generate a high number of centroids that 
represent underlying data points with finer detail. Using a very large number 
of centroids, our indexing method can be applied to any metric space, 
including the particular case of vector spaces. We show that this unusual 
application of $k$-means has attractive properties and renders good performance,
even with high-dimensional datasets, represented in a feature space that is 
usually sparse.
\smallskip

Scalability is another important limitation for many indexing methods, that 
struggle to work with large datasets \citep{Dohnal2008}. Indexes whose
design follows a top-down data partition strategy are not prepared for big 
data problems in distributed systems, where it is unfeasible to centralize 
metadata in a single node. Instead, successful indexing methods for large 
datasets usually follow a bottom-up partitioning strategy. We propose the same 
approach in MASK, which makes it suitable for 
distributed computing applications. In this case, an independent, multilevel 
index structure can be created for each data partition without the need of any
information exchange between computing nodes, either to build the indexes or
to solve queries. At search time, the top level index at each node can  
discard partitions that cannot contain a valid answer, to speed up the 
retrieval of candidate elements.
\smallskip

The rest of the paper is organized as follows.
Section~\ref{sec:related-work} reviews previous research related
to multidimensional data indexing, in particular MAM and,
within them, those based on clustering algorithms. Section~\ref{sec:multilevel-k-means}
describes MASK, the new indexing method based on an
unconventional application of $k$-means. Section~\ref{sec:evaluation} presents
the results of two empirical experiments to gain intuition about the
properties exploited by this new indexing method, as well as to evaluate its
performance in high-dimensional, high-sparsity problems. In Section
~\ref{sec:discussion} we discuss further design considerations
and possible applications of this indexing method, including its application
for similarity search in distributed computing platforms. Finally, in
Section~\ref{sec:conclusion} we recap the main conclusions for this work
and lay out promising lines for future work.

\subsection{Contributions}
\label{subsec:contributions}

MASK follows a multilevel, bottom-up design, similar 
to several methods found in previous research (see Section
\ref{sec:related-work} for further details). However, unlike extant 
proposals for approximate indexing, it is based on an unconventional application 
of the $k$-means algorithm, that can take advantage of increasing resources 
in modern computing systems and distributed platforms. Following this 
strategy, data points can be assigned to closer $k$-means centroids, which are
distributed in the feature space according to the density of data in different 
regions. Thanks to this, MASK can improve the performance of similarity 
search queries in high-dimensional problems which, oftentimes, also exhibit 
high sparsity. In consequence, it is characterized by two distinctive traits:

\begin{itemize}
    \item The multilevel structure of clusters is built following a bottom-up 
    approach. As a result, index construction can be distributed among 
    several computing nodes, where each node can store one or several data partitions. 
    Furthermore, we show that this strategy let the index recover underlying 
    patterns in the dataset represented, in the general case, via pairwise 
    dissimilarities between elements that define a dissimilarity space~\citep{Duin2012}.
    \smallskip
    \item Rather than determining a restrained number of prototype
    points ($k$-means centroids) to cluster elements at each level, our 
    indexing method \textit{bombards} each data partition with \textit{as many 
    centroids as it can be afforded} by available resources in each computing 
    node. Once all centroids have been placed at their final location, they 
    become the new set of elements to be clustered at the next layer above, 
    using again as many $k$-means centroids as possible. This procedure is 
    repeated at each additional level aggregated to the index structure, until 
    the set of centroids at the final top level becomes of manageable size.
\end{itemize}


\section{Background and related work}
\label{sec:related-work}

This section recaps the main concepts and strategies that constitute 
the basis for different similarity search methods. Featured access methods related 
to MASK are also described. Our main goal is to contextualize our work with 
previous research in this area and highlight the main novelties introduced 
by MASK, with respect to previous approaches.

\subsection{Exact similarity search}
\label{subsec:exact-simsearch}

As described above, access methods for exact similarity search include SAM
(Spatial Access Methods) and MAM (Metric Access methods). Regarding SAM, surveys 
comparing dozens of indexing algorithms in this class can be found~\citep{Gaede1998,Bohm2001}. 
Archetypal examples include the B-Tree~\citep{Bayer2002btree, Comer1979}, Bloom 
filters~ \citep{Bloom1970, Alexiou2013bloom}, the k-d tree~\citep{Bentley1975, Friedman1977}, 
linear quadtrees~\citep{Finkel1974}, as well as the R-tree~\citep{Guttman1984} and 
its variants.

Except for few exceptions, these methods render very poor performance
partitioning high-dimensional representation spaces. Experiments conducted by 
\cite{Nene1997} suggest that they become unusable for a number of dimensions 
greater than 15. In the same way, according to \cite{Beyer1999} problems start 
at 10-15 dimensions in experiments with data following a uniform distribution.
In that case, or when there is not even a clear representation of the elements
in the set and only a distance function can be used to compare any pair of
them, it is necessary to resort to MAM.
\smallskip

Different MAM have been proposed to implement indexing structures and
query operations in metric spaces. Some of them use clustering algorithms, like $k$-means 
\citep{Steinhaus1956,MacQueen1967,Lloyd1982}, to recursively partition the
whole set of elements, leading to the creation of a hierarchical structure (tree)
of nested clusters. This tree of clusters is usually built following 
a top-down approach. Furthermore, in all cases the utilization of the $k$-means 
algorithm in index creation follows the standard procedure of estimating
a restricted number of centroids, around which nested subgroups of elements can be
grouped at each level.


As in the case of SAM, several surveys \citep{Chavez2001, Hjaltason2003, Hetland2009}
and comprehensive monographs ~\citep{Samet2006, Zezula2006} provide thorough
comparisons among alternative MAM and their properties.\cite{Uhlmann1991} introduces
general properties of these indexing structures, along with an initial 
taxonomy to classify MAM according to two possible data partition strategies:

\begin{itemize}
    \item \textit{Ball partitioning}: In this type of indexing methods a
    subset of featured elements (sometimes referred to as \textit{vantage points})
    $\{v_i\} \in \mathcal{S},\; i=1,\dots,n$ 
    are selected and a ball of radius $r$ is defined around 
    each of them, with $r = \mathrm{median}(d(v_i,a)),\; \forall a \in \mathcal{S}$. 
    As a result, each ball defines a data partition
    separating all points that lie within the scope of the ball from all
    other points outside the ball. For a dataset distributed uniformly
    over the metric space this method is known to create many partitions
    that will be intersected by the query region in many similarity search
    problems, thus providing poor performance~\citep{Uhlmann1991, Brin1995}.
    \smallskip
    
    \item \textit{Generalized hyperplane partitioning}: This class of MAM relies
    on a data partitioning scheme based on defining a set of generalized
    hyperplanes (GH). Given two elements $v_1, v_2 \in \mathcal{S}\; |\; v_1 \neq v_2$
    a GH is defined as the subset of elements 
    $\{q_i\} \in \mathcal{S} \; |\; d(q_i,v_1) = d(q_i, v_2),\; \forall i$.
    In consequence, each GH partitions the space in two regions, one for all
    elements closer to $v_1$ and the rest of elements closer to $v_2$. This
    strategy can produce more balanced partitioning schemes with GH
    defined using randomly sampled elements.
\end{itemize}

The Vornoi-Tree \citep{Dehne1987} is a well-known example of a ball
partitioning indexing algorithm. In the same way, one of the most popular
algorithms for indexing in metric spaces, the M-Tree \citep{Ciaccia1997},
is also a prominent example of a ball partitioning algorithm (although
it partially incorporates some aspects of GH methods). In contrast,
the generalized hyperplane tree (GHT) \citep{Uhlmann1991} and its
extension into an \textit{m}-ary tree, the GNAT \citep{Brin1995} are
two featured examples of GH-based indexing methods.

Complementing the previous taxonomy, \cite{Chavez2001} present a coherent
framework to analyze data indexing methods in metric spaces, mainly from
the point of view of multimedia databases and information retrieval systems,
and introduce an alternative classification of MAM in two groups:

\begin{itemize}
    \item \textit{Pivoting algorithms}: A subset of reference elements
    $\{v_i\} \in \mathcal{S},\; i=1,\dots,n$ (known as \textit{pivots})
    is identified. Then, all remaining objects are classified according
    to their distances to the pivot objects. Clearly, ball partitioning
    methods fall in this category, which also includes other MAM that
    make use of precomputed distance matrices between pairs of elements in the
    dataset, such as AESA~\citep{VidalRuiz1986} and LAESA~\citep{Mico1994}.
    AESA is very fast but it consumes a lot of resources (requires $O(n^2)$ 
    space and construction time). LAESA uses $k$ fixed pivots so that
    required space and construction time is reduced to $O(kn)$.
    \smallskip
    
    \item \textit{Compact partitioning algorithms}: In this case, the
    metric space is partitioned into clusters, according to the proximity
    of elements to the centroids of each cluster. In this case, it is
    guaranteed that each element is associated to its closest cluster
    center, whereas in pivoting algorithms that may not be the case
    for elements associated to certain pivots. Methods based on the
    GH pertain to this category, as well as algorithms defining a
    tree of nested clusters.
\end{itemize}

The Burkhard-Keller Tree \citep{Burkhard1973} is probably one 
of the first examples of pivot-based indexing algorithm, although it is 
conceived only for discrete distances. \cite{Uhlmann1991} introduces
de \textit{metric tree}, whose design is further expanded in the
VP-Tree \citep{Yianilos1993} and introduces the alternative term
\textit{vantage point} to name the pivots. In contrast,
the hierarchical $k$-means tree \citep{Fukunaga1975} is one of the first
examples of compact partitioning algorithm. Besides GHT and
GNAT, another example of compact partitioning algorithm is the
list of clusters (LC) \citep{Chavez2005LoC}. This algorithm
exhibits very good performance for indexing high-dimensional spaces,
but at the cost of a quadratic complexity in construction time,
which invalidates it for distributed computing and big data settings.
\smallskip

Most of these indexing methods are designed for static datasets, and
they are not well prepared for frequent insertions and deletions.
Recent variants of MAM are specifically conceived for dynamic data, such 
as the DBM-Tree~\citep{Vieira2010}. Its index structure is adapted to the 
density of local data regions, so that the height of the tree is higher 
in denser areas to attain a trade-off solution. Moreover, other recent proposals 
for MAM also include: probabilistic approximate indexing~\citep{Murakami2013}, 
which sacrifices accuracy for query resolution speed (see Section 
\ref{subsec:approx-simsearch} below); redesigns of classic indexes such as 
the M-Tree, using cut-regions instead of ball regions~\citep{Lokoc2014cutreg}; 
specific methods to improve in-memory data indexing~\citep{Carelo2011,Pola2014}; 
a disk-based method to be integrated in commercial database management systems
(SPB-Tree~\citep{Chen2017spbtree}) and new approaches to increase efficiency 
in dynamic data indexing~\citep{Oliveira2017}.

\subsection{Approximate similarity search}
\label{subsec:approx-simsearch}

The approach proposed in this paper to resolve similarity queries pertains
to the class of approximate search methods. Many examples of this
kind of algorithms can be found in previous research. They usually aim at
situations in which exact methods cannot provide a fast answer, such as
in high-dimensional problems or using big data, where exact techniques 
default to a full scan of the entire dataset. 

It is possible to find approximate similarity search
methods designed for either vector spaces or 
metric spaces. \cite{Skopal2007appsearch} introduces a unified framework
to study fast similarity search methods that can be either exact or approximate.
Besides, \cite{Patella2009appsearch} present a comparative summary of
relevant approximate methods, introducing a taxonomy to classify 
them according to relevant traits, such as target
space (vector, metric), strategy to obtain approximate results
(changing the representation space, reducing the number of comparisons),
guarantees provided on the quality of results (no guarantees, deterministic,
probabilistic) and degree of interaction with users (static, interactive). 
\smallskip

One of the first algorithms for approximate similarity search in metric spaces 
found in literature is FastMap \citep{Faloutsos1995FastMap}. It is based on
a method to obtain a fast projection of elements in the original dataset into
a $k$-dimensional space, where $k$ is a user-defined parameter and the projection
preserves the distance between pairs of elements. Thus, it can reach an
approximate answer via dimensionality reduction. Other popular exact
algorithms also have their approximate versions, including three variants
for the R-Tree \citep{Corral2005} in vector spaces and three more based
on the M-Tree \citep{Zezula1998appmtree} in metric spaces.

Certain approximate algorithms combine different strategies. An interesting
example is the Integrated Progressive Search \citep{Ferhatosmanoglu2001},
which aims at vector spaces. This method first applies dimensionality
reduction methods and, later on, applies conventional $k$-means clustering
\citep{Xu2009clustering} on the resulting data points in a lower 
dimension space. This approach reduces the number of distance 
comparisons that must be performed. However, in general it is not straightforward 
to find a dimensionality reduction technique that works well for any problem. 
As we will see later, MASK takes a different approach, leveraging some interesting
properties of the core $k$-means algorithm \citep{Arthur2007} to reach an 
approximate result without the need to reduce the number of dimensions of the 
original feature space.
\smallskip

Approximate similarity search methods have become a frequent
approach applied in previous research works, since they provide
a convenient trade-off between speed and accuracy of results. Featured
methods include Locality Sensitive Hashing (LSH) \citep{Gionis1999}, which
has interesting applications for similarity search under edit distance
\citep{McCauley2021}, and the DAHC-Tree \citep{Almeida2010}, aimed at
high-dimensional problems in metric spaces. The indexing method proposed
in this paper takes advantage of certain properties of the $k$-means algorithm,
along with available computational resources, to provide an approximate
indexing method that is easy to construct, valid for the general case
of metric spaces and can be used with both static and dynamic datasets.
Therefore, it offers a versatile compromise solution for similarity search.

\subsection{Scalable similarity search}
\label{subsec:distrib-simsearch}

Several indexing methods for similarity search has been designed
to work with distributed systems~\citep{Hetland2009}, addressing large
and complex datasets. Some initial examples of this kind are GHT$^*$~\citep{Batko2004}, MCAN~\citep{Falchi2007mcan}, Metric Chord (M-Chord)~\citep{Novak2006mchord} and an extension of GNAT (EGNAT) 
for parallel systems~\citep{Marin2007egnat}. 

Later on, implementations in real systems started to appear,
such as MD-HBase~\citep{Nishimura2013}. This method
integrates multidimensional indexing in the open source
key-value store HBase, focusing on location based services.
Nevertheless, this application only implements two examples of SAM (K-d Tree
and the Quad Tree) in the distributed data store, without considering
metric spaces. Likewise, the A-Tree~\citep{Papadopoulos2011} implements
a combination of R-trees and Bloom filters (both part of the SAM family) for 
multidimensional data indexing in cloud computing platforms. M-Grid
\citep{Kumar2017} is a more recent example of a distributed multidimensional
indexing structure for location based data, that claims to substantially 
improve the performance of MD-HBase. Another innovative
example in the MAM family is D-Cache~\citep{Skopal2012dcache}. This method
is based on caching distance information that can be leveraged in 
distributed systems to speed up similarity queries.
\smallskip

Nowadays, distributed systems involving big data and high-dimensional, high-sparsity problems have proliferated. 
Technological frameworks like Apache Spark
have been determinant for the widespread adoption of cluster-based
systems in many areas, including machine learning. Recent advances show 
promising implementations of SAM on Apache Spark, specifically for 
similarity search with spatial data and IoT applicatons
\citep{Limkar2019,Elmeiligy2021}. However, to date there is no clear
implementation of more general MAM in modern distributed data processing
frameworks. As we describe in Section~\ref{sec:multilevel-k-means},
the approximate similarity search method proposed in this paper can
be naturally implemented on distributed computing platforms. The 
multilevel, bottom-up approach to build the index can
be independently executed in each data partition, located in different
nodes. Furthermore, the index construction does not
imply any exchange of information or metadata between nodes storing different
data partitions. The only coordination requirement would be to maintain
the top level layer of representative points, to help in deciding which
nodes should be involved in resolving a particular query.

\subsection{Machine learning and similarity search}
\label{subsec:ml-simsearch}

As we have seen, there is a close connection
between machine learning and similarity search indexing. The foundations
for designing data access methods repose on familiar concepts
in machine learning classification, such as dissimilarity spaces and 
distance functions. They are also affected by the same limitations,
namely the curse of dimensionality and the challenge of identifying the actual
intrinsic dimension governing some problems.

Clustering algorithms play a central role in many indexing methods for
similarity search. The $k$-means algorithm~\citep{Steinhaus1956,MacQueen1967} 
stands out as a recurrent solution to build indexing structures, starting
with hierarchical $k$-means~\citep{Fukunaga1975}. In fact, according to
\cite{Samet2006}, the GHT and GNAT indexing methods can be regarded as
special cases of a broader class of hierarchical clustering methods
described by \cite{Burkhard1973} and \cite{Fukunaga1975}. However,
the pivots selected by GNAT are not necessarily $k$-means centroids, which
are $k$ points that may not belong to $\mathcal{S}$ and that minimize the
sum of squared distances of individual objects to their closest centroid.

Interestingly, the applicability of $k$-means for the construction of
hierarchical indexing structures is already described, albeit without 
implementation details, by \cite{MacQueen1967}. 
This application entails building tree clusters in a top-down fashion, 
so that the within cluster variance does not exceed an upper
threshold $R$. The set of centroids at each level act as a fair representation
of data objects, effectively summarizing the clusters of all lower
levels. Likewise, MacQueen also demonstrates that $k$-means can be 
readily extended beyond vector spaces to the general case of metric spaces. 
For example, $k$-means has been successfully applied in image
indexing~\citep{Cao2013infoKmeans}. A recent application of 
$k$-means and Voronoi diagrams for multidimensional data indexing and 
spatial data query processing in sensor  networks~\citep{Wan2019kmeans} 
confirms the validity of multilevel $k$-means indexing for contemporary 
distributed data problems. 

Nevertheless, this approach for building hierarchical indexes
with $k$-means clusters is based on establishing an upper limit $R$ for
within-cluster variance. In fact, this is just an alternative formulation
of the typical problem in partitioning clustering, namely selecting
the appropriate number of clusters that must be identified in the dataset
\citep{Gordon1999classification, Aggarwal2015dm}. Many different methods
have been proposed to solve this problem (see, for instance,
\citep{Milligan1985, Charrad2014} for a detailed discussion), some of which
rely on fixing a within-cluster variability threshold 
\citep{Mirkin2005clustering,Xu2009clustering}.

To avoid these practical issues, it would be desirable to just let the
multilevel clustering algorithm to adapt itself to the density of 
objects in different regions of the metric space to be indexed. In 
a certain way, this resembles the idea behind the DBM-Tree discussed 
in Section~\ref{subsec:exact-simsearch}, but considering in this case 
a tree of clusters. Our indexing algorithm introduces an
unconventional application of $k$-means because it completely eliminates
the restriction of determining the optimal number of clusters in advance.
Instead, we propose to create as many centroids at each level as it can 
be afforded by our computational resources. 

As we show in Section~\ref{sec:multilevel-k-means}, we can 
take advantage of the minimization of the sum of square distances from 
each object to the closest centroid to adjust the position of prototypes
for each cluster, according to the density of elements in each region
of the metric space. In a certain way, this approach also follows the recent
strategy of \textit{learned indexes} in databases~\citep{Kraska2018learnedidx},
which states that traditional indexing structures can be substituted by
multiple levels of machine learning algorithms that can help to approximate
the underlying distribution of data. The Z-order Model (ZM) index is
an example application of the learned index strategy to spatial index
structures (for instance, the R-tree). In our method, we let the $k$-means
algorithm to learn good locations to place the centroids representing
each cluster, assuming that a sufficiently large number of centroids is
used to map the set $\mathcal{S}$.



\section{Multilevel k-means structure for data indexing}
\label{sec:multilevel-k-means}

This section describes the MASK method for approximate similarity search in 
metric spaces, using a multilevel index structure. The rationale behind the 
unconventional application of the $k$-means clustering algorithm is also
explained. This method can be implemented in distributed systems, as different 
sections of the multilevel index can be independently created on data partitions 
stored in different nodes.

\subsection{Multilevel index design}
\label{subsec:design-overview}

\begin{figure}[ht!]
    \centering
    \includegraphics[width=0.9\textwidth]{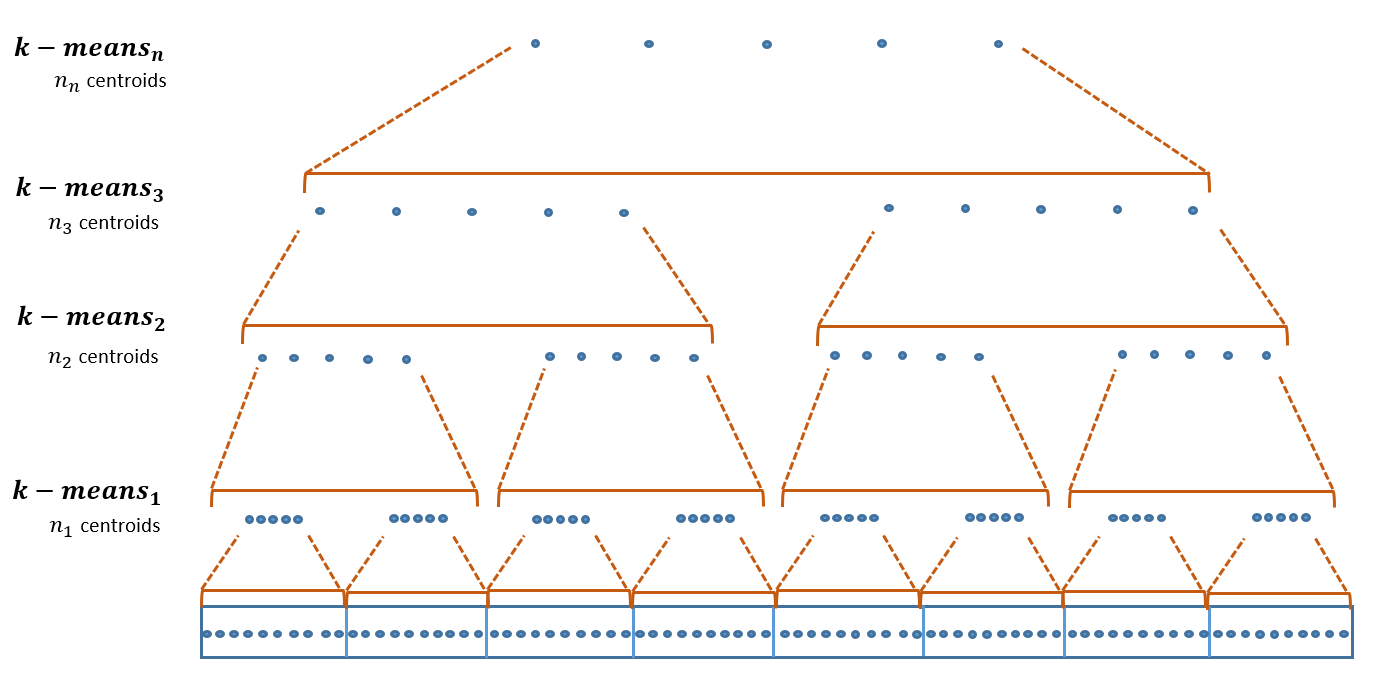}
    \caption{Conceptual representation of the multilevel structure approach followed in MASK.}
    \label{figure-multilayer}
\end{figure}

Figure~\ref{figure-multilayer} represents the design principles of MASK. At the lowest 
level, we have the data points in $\mathcal{X}$ to be indexed. Blue boxes in the diagram 
represent different data partitions or groups. These groups could be stored in different 
nodes of a cluster although, for simplicity, in this case data partitions are considered to 
reside in the same node.

As described in Section \ref{sec:related-work} above, several MAM rely on building a 
hierarchy of clusters, using unsupervised machine learning algorithms like $k$-means. In
the case of MASK, the multilevel structure of $k$-means centroids is created
following a bottom-up approach. The first level of centroids (identified as $n_1$ in
Figure \ref{figure-multilayer}) summarize the actual data points at the lowest level,
so that the points assigned to each cluster are represented by its $k$-means 
prototype (centroid). The second level (identified as $n_2$), corresponds to centroids 
representing the prototypes calculated at the first level, $n_1$. The same procedure is 
recurrently applied, creating additional layers of prototypes summarizing the immediately 
lower group of centroids. The method stops when the number of centroids at the top layer of 
this multilevel index structure becomes manageable. Algorithm \ref{algo:multilevel_structure}
details the construction of this multilevel index structure, using $k$-means.

\begin{algorithm}
    \SetKwFunction{split}{split}
    \SetKwFunction{kmeans}{kmeans}
    \KwIn{data, lengthGroup, nCentroids}
    \KwOut{layerPoints, layerLabels, layers}
    \BlankLine
    $ngroups \leftarrow lengthData/lengthGroup$\;
    $vector \leftarrow$ \split{$data, ngroups$}\;
    Initialize lists $layerPoints$ and $layerLabels$, and variable $idLayer$\;
    \While{$ngroups \geq 1$}{
        \tcp{Initialize groupsPoints, groupsLabels and points}
        \For{$idGroup=1$ \KwTo $ngroups$}{
            $points \leftarrow vector[idGroup]$\;
            $groupsPoints.add($\kmeans{$nCentroids, points$})\;
            $groupsLabels.add($\kmeans.labels)\;
        }
        $layerPoints.add(groupsPoints)$\;
        $layerLabels.add(groupsLabels)$\;
        $ngroups\leftarrow length(groupsPoints)/lengthGroup$\;
        \If{$ngroups \geq 1$}{
            $vector \leftarrow$ \split{$layerPoints, ngroups$}\;
        }
        $idLayer \mathrel{+}= 1$\;
    }
    $layers\leftarrow idLayer-1$
    \caption{Multilevel index construction}
    \label{algo:multilevel_structure}
\end{algorithm}

The assembly process requires two initial parameters: the group length ($lengthGroup$) 
and the number of centroids calculated for each group ($nCentroids$). 
Group length refers to the number of elements in each data partition at the lowest
level. In general, we assume that all partitions will have the same size, although this
is condition is not strictly necessary. The size of partitions should be configured 
according to available computing power (and the total number of nodes in distributed 
computing systems).

Figure \ref{figure-multilayer} shows a simple case example: 80 points in the
dataset at the lowest level, with $lengthGroup=10$ and $nCentroids=5$. Thus, at the 
first level the algorithm assumes 8 groups of 10 points. In each group, the algorithm 
builds 5 clusters using $k$-means. Depending on the distance between points and their
corresponding centroid, a different number of points can be assigned to each cluster.
For this reason, regions with higher data density will also tend to receive a higher
number of centroids. This clustering at the first level reduces the initial number of data 
points by a certain proportion, the \textit{data summarization ratio}, determined
by the relationship between the values of $lengthGroup$ and $nCentroids$. In this
example, this ratio is 2:1 and, hence, the size of each group of data is shrunk by a 
half, from 80 to 40. 

At the second level, the algorithm takes 4 groups of 10 points (since $lengthGroup=10$)
and $k$-means is applied in each group to obtain 5 new clusters, each one represented
by a centroid. This process is repeated recursively, until the number of points at the
top level is small enough. The stop condition implemented in this case is that
the total number of centroids at the top level must be less than or equal to 
$lengthGroup$. Therefore, the computational complexity of the problem gets lower as
the total number of points is reduced after each iteration.
\smallskip

Nevertheless, the important aspect of selecting the optimal number of centroids to 
calculate at each level remains open. In general, previous indexes based on trees of 
clusters generate a limited number of centroids at each level. In contrast, recent 
methods for metric spaces, such as DBM-Tree~\citep{Vieira2010}, adapt the height of 
the indexing tree to changes in data density of different regions in 
$\mathcal{X}$. However, the downside of the latter approach is that we may not attain 
uniform search performance, since the algorithm must traverse more or less levels
depending on the density of the target region for each query. To solve this issue, the 
core novelty in MASK is an unconventional application of the $k$-means algorithm, 
tailored to the specific case of information indexing in modern computing platforms, 
that departs from the traditional strategy suggested in clustering problems.

\subsection{Using k-means for information indexing}
\label{subsec:k-means-index}

Typical usage of $k$-means in unsupervised machine learning dictates that the appropriate 
number of clusters for a given problem must be found beforehand, where each cluster is 
represented by a centroid. Therefore, the number of centroids should not be too large, 
since they would not summarize underlying data effectively, or too small, which would 
group unrelated data points together. In practice, the optimal number of centroids can 
be determined in different ways, such as using the \textit{elbow} method \citep{Watt2020}, 
based on a plot of the total within-cluster sum of squares for different $k$, the 
Silhouette Coefficient Algorithm \citep{Kaufman1990} or the Gap statistic 
\citep{Tibshirani2001}. Moreover, the final result can be quite sensible to the choice of
initial locations for cluster centroids. To circumvent this limitation, algorithms such 
as $k$-means++ \citep{Arthur2007} propose spreading the initial random centroids more 
evenly which, generally, leads to better results.
\smallskip

However, the goal pursued by MASK is approximate data indexing, not clustering. 
As more powerful computational infrastructures become available, providing larger 
memory and storage capacity, prior restrictions about the number of clusters to maintain 
at each level become less relevant. In this new scenario, it is interesting to check 
what happens when the dataset is \textit{bombarded} with a very high number of $k$-means 
centroids, as large as it can be afforded by the computing infrastructure capacity.
\medskip

A conceptual experiment can be useful to illustrate this approach, on a dataset comprising 
4 clouds generated from a  t-distribution in $\mathbb{R}^2$, with 12 degrees of freedom 
and different means for each cloud, so that they  are clearly separated from each other. 
The experiment consists of two tests:

\begin{itemize}
    \item In the first test, an increasing number of $k$-means centroids (from 4 to
    128) are generated on the complete dataset, following a top-down approach. That is, 
    the complete set of points is provided to $k$-means to calculate the position 
    of centroids in each case.
    
    \item In the second test, data points are first randomly assigned to 4 different
    data partitions (groups), simulating the situation that MASK would find in a 
    distributed system. Then, each group is independently bombarded with a growing
    number of $k$-means centroids, so that, for each case, the aggregated number of
    centroids in all groups is the same as in the first test. This represents 
    a bottom-up approach for data indexing.
    
\end{itemize}

\begin{figure}[ht!]
    \centering
    \includegraphics[width=0.9\textwidth]{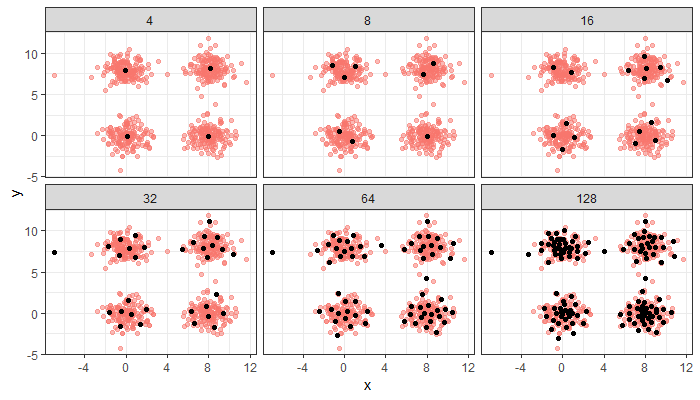}
    \caption{Results of bombarding 4 data clouds with an increasing 
    number of $k$-means centroids, following a top-down approach. Labels on each pane indicate the total number of centroids generated for each case.}
    \label{figure-topdown}
\end{figure}

The experiment aims to demonstrate that quite similar results can be attained disregarding
the approach adopted for data indexing. Figure~\ref{figure-topdown} shows the result of 
the first test. All panels depict the same 4 t-distribution clouds, which are
bombarded with an increasing  number of $k$-means centroids (in black). 
Interestingly, we observe that, as the number of centroids increases, more of them 
tend to concentrate in denser data regions. This behavior is in line with existing results on the consistency of the k-means method~\citep{Pollard1981}. When using a very large number of centroids,
some of them even gravitate towards outliers, providing effective coverage also for 
extreme data points.

\begin{figure}[ht!]
     \centering
     \subfloat[Partial results of generating 8 centroids (in black) within each data partition (group). Data points from the 4 original clouds are randomly assigned to each partition.\label{fig:data-partitions}]
         {\includegraphics[width=0.8\textwidth]{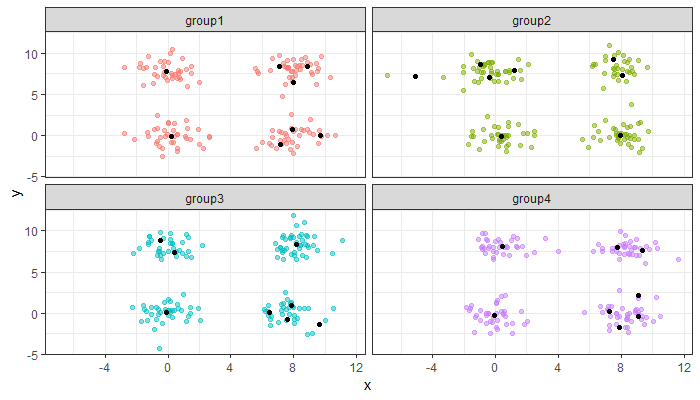}}
     
     \subfloat[Aggregation of partial results from (a) in a single plot. Data points are coloured according to their data partition. A total of 32 centroids (in black) are created. \label{fig:centroids8}]
         {\includegraphics[width=0.8\textwidth]{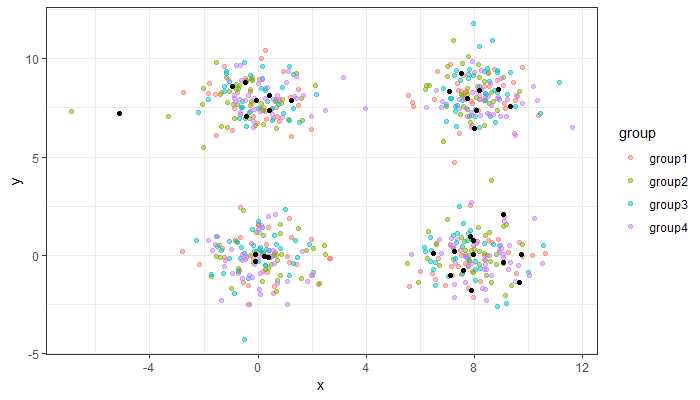}}
    \caption{Indexing in the second test, using 32 $k$-means centroids in total (8 centroids per group). Partial results in (a) are combined in a single plot in (b) for comparison with the equivalent case in the first test.}
    \label{figure-explanation-bottomup}
\end{figure}

Figure~\ref{figure-explanation-bottomup} represents the bottom-up indexing procedure in 
the second test, and how partial outcomes from each group are combined to report
the final result, so that each case can be compared with its counterpart in the first
experiment.

\begin{figure}[ht!]
    \centering
    \includegraphics[width=0.95\textwidth]{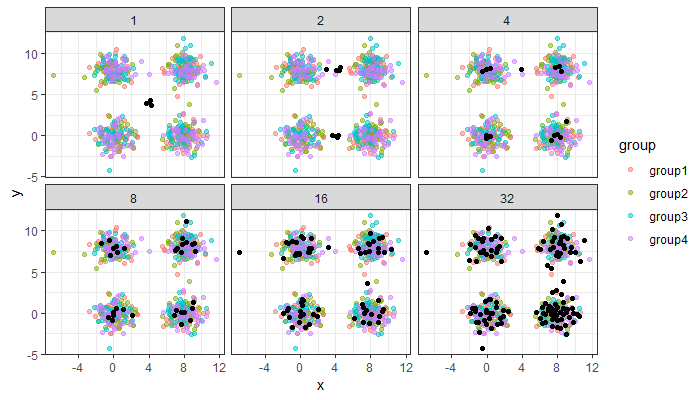}
    \caption{Results of bombarding the same dataset with an increasing number of $k$-means centroids, following a bottom-up approach with separate data partitions. Points are coloured according to the data partition (group) to which they have been assigned. Labels on each pane indicate the number of centroids generated within each group.}
    \label{figure-bottomup}
\end{figure} 

Figure~\ref{figure-bottomup} presents the aggregated results for each case in the 
second test. As explained in Figure~\ref{figure-explanation-bottomup}, each group can contain points that belong to any of the 4 initial clouds. Finally, 
the $k$-means algorithm is run independently on each group, using a growing 
value of $nCentroids$, from 1 to 32. Each pane in Figure~\ref{figure-bottomup} plots 
the aggregation of all partial results. Labels on top of each pane indicate the value of 
$nCentroids$ calculated in each group, so that the total number of centroids for each 
case is the same as in the previous experiment. Results from this second test are 
comparable to those presented in Figure~\ref{figure-topdown} above, provided that a high
enough number of centroids is used for data indexing. This demonstrates that data 
partitions can be stored in separate nodes and, then, independent indexing structures 
can be created within each node. Hence, MASK can scale up to handle problems involving 
big data distributed in parallel computing clusters.

\subsection{Searching and data insertion}
\label{subsubsec:search-algo}

When the index based on the multilevel structure with $k$-means centroids
is ready, we can use it to accelerate similarity search queries. The
hierarchy of centroids can help us discarding sections of the dataset where
it is unlikely to find candidate results. However, the $k$-means
algorithm usually provides a local optimum solution and cannot warrant
to find the global optimum for the location of centroids. In consequence, 
we cannot assure that the search process using our multilevel index
returns exact results. 

For this reason, our algorithm pertains to the class of approximate 
similarity search methods. In spite of this apparent limitation, we will
show that time and resources required to build the index are affordable and,
depending on the characteristics of the dataset being indexed, the accuracy
of results can be sufficiently high for many practical applications.

To solve any type of query using this multilevel index, we start at the 
top level of the hierarchy of centroids, steering the search according to the 
distance between the centroids at each level and the target query object 
$q \in \mathcal{X}$. Then, we can adapt this general approach to solve a 
specific type of query, as follows.

\begin{figure}[ht!]
    \centering
    \includegraphics[width=0.9\textwidth]{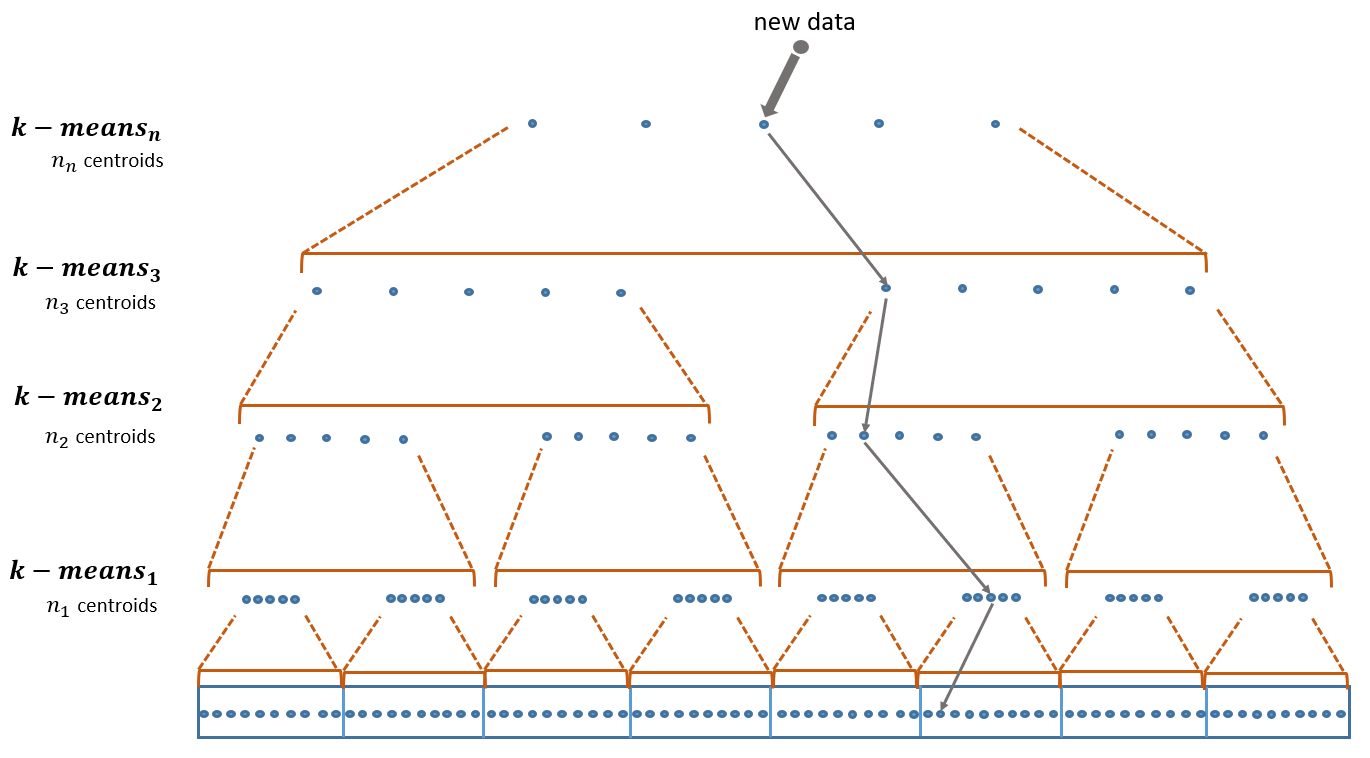}
    \caption{Procedure to perform a point search using the proposed multilevel
    index based on $k$-means.}
    \label{figure-search}
\end{figure}

    \paragraph{Point query} At the top level, the distance from each 
    centroid to $q$ is calculated, and the centroid with the minimum 
    distance value is selected. Then, at the next level only the subset of 
    centroids represented by the one previously selected
    at the top level is considered. Again, the distance from each centroid in 
    this subset to $q$ is calculated and the centroid with the minimum
    distance is chosen. Subsequent iterations repeat the same procedure at
    each layer over the remaining levels, (focusing on children of the selected
    centroid at the previous level and taking the child with the minimum
    distance to $q$). Eventually, the algorithm reaches the subset of 
    actual data points  represented by the centroid selected at the 
    lowest layer of the index.
    At this stage, it is easy to perform a full scan over this small subset
    of data points and return the one with the lowest distance to $q$ as
    the result of the point query. Figure \ref{figure-search} illustrates
    this process, marking with arrows the centroid selected at each level
    of the index. This procedure is almost equivalent to the nearest-neighbour 
    search described in Algorithm \ref{algo:nn-search}, below. The only 
    difference is that, here, the algorithm seeks a perfect match within the 
    data partition reached at the end of the search process.

\paragraph{Nearest neighbour and $k$-nearest neighbours queries} The
    search for the nearest-neighbour can be performed in exactly the same
    way as to resolve the point query. However, instead of looking for an
    exact match, in this case the element with minimum distance to $q$ found 
    within the data partition is returned. If the query searches for the $k$ 
    closest elements to $q$, the final subset of data points is ranked according 
    to their distance to $q$, and the query returns the $k$ members with the 
    lowest distance values from the rank. Algorithm \ref{algo:nn-search} illustrates 
    the case of the k-nearest-neighbour search.

\begin{algorithm}
    \SetKwFunction{euclideanDistance}{euclideanDistance}
    \SetKwFunction{searchPosMin}{searchPosMin}
    \SetKwFunction{searchKNN}{searchKNN}
    \SetKwFunction{searchGroup}{searchGroup}
    \KwIn{layerPoints, spoint, layers, nCentroids}
    \BlankLine
    \For{$idLayer=layers$ \KwTo $1$}{
        \If{$idLayer \neq layers$}{$idGroup \leftarrow$ \searchGroup{$layerLabels, idLayer$}}
        \Else{$idGroup \leftarrow 0$}
        $centroids \leftarrow layerPoints[idLayer][idGroup]$\;
        $matrixD \leftarrow$ \euclideanDistance{$spoint, centroids$}\;
        $posCentroid \leftarrow$ \searchPosMin{$matrixD$}\;
        \tcp{Correction of the centroid identifier}
        \If{$idLayer \neq layers$}{
            $idGroup \leftarrow posCentroid/nCentroids$\;
            $idCentroid \leftarrow posCentroid - (idGroup * nCentroids)$\;
        }
        \Else{
            $idGroup \leftarrow 0$\;
            $idCentroid \leftarrow posCentroid$\;
        }
    }
    \tcp{Data layer}
    $selecPoints \leftarrow layerPoints[idGroup][idCentroid]$\;
    $matrixD \leftarrow$ \euclideanDistance{$spoint, selecPoints$}\;
    $idKPoints \leftarrow$ \searchKNN{$matrixD$,$k$}\;
    \caption{$k$-NN search algorithm.}
    \label{algo:nn-search}
\end{algorithm}

\paragraph{Range query} This type can be addressed considering,
    at each level, the subset of children centroids whose distance to $q$
    lies within a given ball $r$ around $q$. Then, the same search path is followed 
    for every children selected in the lower layer. Finally, a set of candidate 
    groups of points will be selected, and the query returns the aggregated collection 
    of points that matches the query condition in each candidate set. Algorithm
    \ref{algo:range-search} details this approach.

\begin{algorithm}
    \SetKwFunction{euclideanDistance}{euclideanDistance}
    \SetKwFunction{sortDist}{sortDist}
    \SetKwFunction{selectCandidates}{selectCandidates}
    \SetKwFunction{searchGroup}{searchGroup}
    \KwIn{layerPoints, spoint, layers, nCentroids, radius}
    \BlankLine
    \tcp{Top layer}
    $idGroup \leftarrow 0$\;
    $centroids \leftarrow layerPoints[layers][idGroup]$\;
    $matrixD \leftarrow$ \euclideanDistance{$spoint, centroids$}\;
    $sortVecCentroids \leftarrow$ \sortDist{$matrixD$}\;
    $vecCandidates \leftarrow$ \selectCandidates{$sortVecCentroids, radius$}\;
    \For{$idLayer=layers-1$ \KwTo $1$}{
        \tcp{Initialize newCandidates, vecChildren}
        \For{$candidate \in vecCandidates$}{
            $idGroup \leftarrow$ \searchGroup{$layerLabels, idLayer$}\;
            $centroids \leftarrow layerPoints[idLayer][idGroup]$\;
            $matrixD \leftarrow$ \euclideanDistance{$spoint, centroids$}\;
            $sortVecCentroids \leftarrow$ \sortDist{$matrixD$}\;
            $vecChildren \leftarrow$ \selectCandidates{$sortVecCentroids, radius$}\;
            \tcp{Correction of the centroid identifier}
            \For{$pos \in vecChildren$}{
                $idGroup \leftarrow pos/nCentroids$\;
                $newCandidates \leftarrow posCentroid - (idGroup * nCentroids)$\;
            }
        }
        $vecCandidates \leftarrow newCandidates$\;
    }
    \tcp{Data layer}
    $matrixD \leftarrow$ \euclideanDistance{$spoint, centroids$}\;
    $sortVecCentroids \leftarrow$ \sortDist{$matrixD$}\;
    $idRangePoints \leftarrow$ \selectCandidates{$sortVecCentroids, radius$}\;
    \caption{Range search algorithm.}
    \label{algo:range-search}
\end{algorithm}


Besides, the MASK method can also be applied to dynamic datasets,
since new points can be stored using the index structure with a 
simple procedure:

\begin{itemize}
    \item First, at the top level select the k-mean prototype with the
    minimum distance to the new point.
    \smallskip
    
    \item Then, from k-mean centroids at the second level
    that are children of the selected prototype at top level, choose again the
    one with the minimum distance to the data point.
    \smallskip
    
    \item Repeat these steps, iterating over the layers of the multilevel
    index until we reach one of the group of data points
    with bounded within-group variance, and assign the new point to 
    that group.
\end{itemize}


It is important to remark that, in case that many new elements are added
to $\mathcal{X}$ following this procedure, the size of data partitions
at the bottom of the hierarchy may become quite uneven. However, this is not
a major problem, since new points are placed next to other
similar elements, guided by the multilayer structure of centroids. 
As a result, if one partition grows beyond a certain threshold, 
it can be split in smaller data groups and reconstruct just the local 
part of the index covering that specific region, without affecting the rest 
of the structure.

\subsection{Indexing distributed datasets}
\label{subsec:distrib-indexing}

\begin{figure}[ht!]
    \centering
    \includegraphics[width=0.9\textwidth]{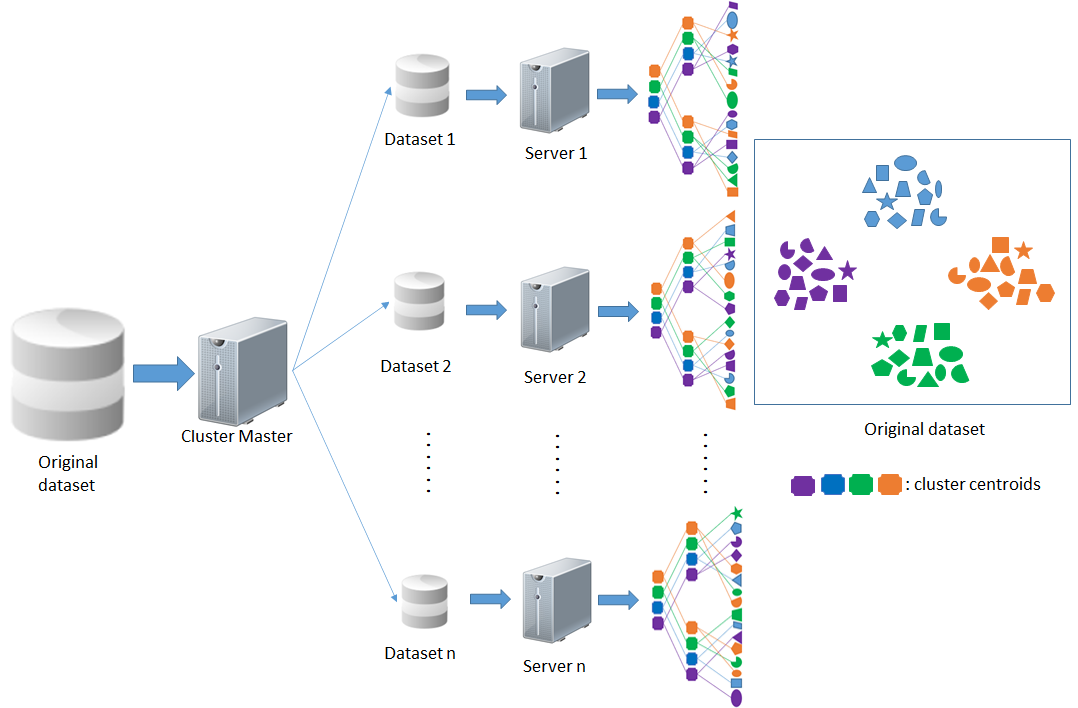}
    \caption{Schematic application of the MASK indexing method on a
    distributed dataset.}
    \label{figure-cluster}
\end{figure}

As noted above, MASK can be implemented in parallel and distributed
computing systems, as shown in Figure \ref{figure-cluster}. In this example,
the dataset consists of four sets of elements, separated among each other 
in the feature space. Besides, let us assume that this dataset is split 
into several partitions, and each partition is stored in a different 
node. Hence, points from any of the four original sets can be found in
any partition. Using MASK, each node can work with its own data 
partitions, in parallel with the rest of nodes. The algorithm is
executed following a bottom-up approach, to obtaining a local multilevel 
index structure in each node. Then, at the management level of the 
distributed system, like the master node of the cluster, just the top 
level centroids from each node need to be recorded. Using these 
metadata, search queries can be launched and MASK can decide which 
nodes will be involved in their resolution.

In the example of Figure \ref{figure-cluster}, MASK creates 
4 centroids at the top level of each node and 2 groups with 4 centroids 
per group at the lower indexing level of each node. Using this distributed
index structure, the algorithm can effectively summarize the original 
dataset, shrinking the size of hierarchical metadata that must be 
maintained by the centralized cluster management service to speed up 
the similarity search. Queries can be solved either searching in all 
nodes in parallel, or restricting the search only to nodes that have
centroids whose distance to the target query object is lower than a 
given threshold $\epsilon$.
\smallskip

Of course, several design parameters can be tuned in each particular
application to adapt the multilevel index construction procedure to
the peculiarities of any specific dataset:

\begin{itemize}
    \item At the lowest level of the index structure, it is important to
    bombard data partitions with as many $k$-means centroids as it can be
    reasonably afforded by available computational resources. As we explained
    in Figure \ref{figure-bottomup}, the higher the number of centroids 
    calculated on the data points, the better will be the accuracy of 
    MASK to find the actual position of points, effectively mapping 
    high-density regions, sparse regions or outliers.
    \medskip
    
    \item At higher layers of the multilevel index, a compromise
    can be attained between indexing accuracy and data summarization. This
    trade-off is determined by the data summarization ratio of centroids 
    between one layer and its adjacent level, below it. For experiments 
    described in Section \ref{sec:evaluation}, we have fixed a value of 2:1
    for the ratio of centroids in adjacent layers,
    effectively halving the number of centroids to be calculated
    in each new iteration. However, further analysis must be conducted
    to evaluate the impact of this configuration parameter on the
    performance of MASK.
    
\end{itemize}


\section{Experimental results}
\label{sec:evaluation}

Several experiments have been conducted to evaluate the performance 
of MASK, using two different datasets:

\begin{itemize}
    \item The first experiment is based on a synthetic dataset including 8 
    Gaussian clouds, generated using the standard procedure described in
    \citep{Jain1988clustering}. In this case, the main goal is to 
    illustrate the behaviour of MASK against a dataset that exhibits 
    well-known theoretical properties.
    \smallskip
    
    \item For the second experiment, we use the Reuters-21578 dataset 
    \citep{Lewis1997},  a popular benchmark in high-dimensional and 
    high-sparsity text classification, obtained from the UCI repository 
    \citep{Dua2017}. Here, we aim to evaluate the capacity of MASK 
    to map and retrieve elements in an adverse scenario for many other
    alternative algorithms.
\end{itemize}

All experiments have been conducted using a server equipped with 2 AMD
EPYC 7451 microprocessors (24 cores/48 threads, 2.30 GHz, 64 MB L3 caché), 
128 GB of DDR4 RAM and an SSD Intel D3-S4510 (capacity 480 GB) for secondary
storage. Next, the methodology developed to undertake these empirical tests 
is described, including the metrics to assess the performance of MASK. 
After this, experimental results are presented.

\subsection{Performance evaluation}
\label{subsection:performance-eval}

The main advantage of MASK is providing a rapid answer to queries, 
at the expense of returning approximate results. Thus, a straightforward 
strategy to assess the performance of this algorithm is to undertake an 
exhaustive search of all elements in a given dataset $\mathcal{X}$, so that 
the proportion of data points that were correctly retrieved can be determined. 
This is the common approach for all evaluation experiments developed in 
this study.

Besides, MASK performance could be improved even more by
relocating points that were not correctly found after the first build
of the multilevel structure of $k$-means centroids. Intuitively, when the
point search reaches a partition at the bottom of the tree and the target 
element is not found in that group, if the target point is reassigned to 
that partition it will join other close elements, according to the distance
function. Then, if the multilevel index is built again it can find 
similar points placed together, making it easier for the point search to
retrieve more elements correctly. In our experiments, several iterations 
of data relocation and index rebuilding are performed, to check whether 
this strategy can help or not to decrease the error rate in some cases.
\smallskip

The main characteristics of the datasets used for performance evaluation experiments 
in this study are summarized in table~\ref{tab:datasets}.

{\renewcommand{\arraystretch}{1.5}
\begin{table}[htb]
    \caption{Notation and description of datasets used to evaluate the performance of the proposed approximate indexing method.\label{tab:datasets}}
    \begin{center}
        \begin{tabular}{ p{1.5cm}  p{4cm}  p{6.5cm} }
            \toprule
            Dataset  & Name & Description \\ \midrule
            $\mathcal{X}_{GNO}$ & Gaussian clouds\newline(no overlap) & 8 Gaussian clouds in $\mathbb{R}^2$, without overlap between adjacent point groups. \\
            $\mathcal{X}_{GMO}$ & Gaussian clouds\newline(moderate overlap) & 8 Gaussian clouds in $\mathbb{R}^2$, with moderate overlap between adjacent point groups. \\
            $\mathcal{X}_{GRO}$ & Gaussian clouds\newline(remarkable overlap) & 8 Gaussian clouds in $\mathbb{R}^2$, with severe overlap between contiguous point groups. \\
            $\mathcal{X}_{REUT}$ & Reuters-21578\newline(UCI repository) & Collection of
            more than 10,000 news documents from Reuters newswire in 1987, representing 90 different to-pics.\\  \bottomrule
        \end{tabular}
    \end{center}
\end{table}
}

\begin{figure}[ht!]
    \centering
    \subfloat[Dataset $\mathcal{X}_{GNO}$ (no overlap).\label{figure_gaussianCloud_noOverlap}]
    {\includegraphics[width=0.5\textwidth]{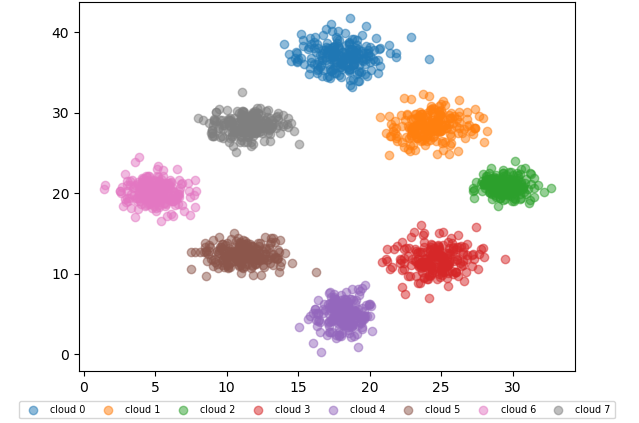}}
    \subfloat[Dataset $\mathcal{X}_{GMO}$ (moderate overlap).\label{figure_gaussianCloud_moderateOverlap}]
    {\includegraphics[width=0.5\textwidth]{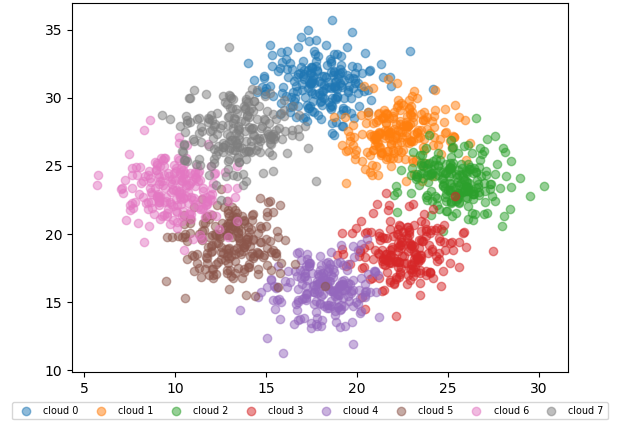}}
    
    \subfloat[Dataset $\mathcal{X}_{GRO}$ (remarkable overlap).\label{figure_gaussianCloud_noticeableOverlap}]
    {\includegraphics[width=0.5\textwidth]{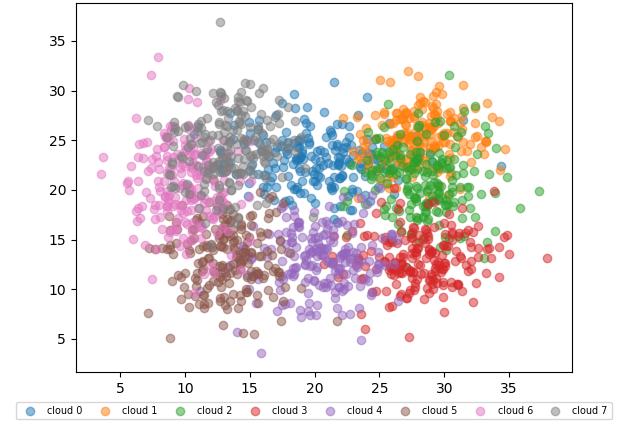}}
    \caption{Synthetic datasets used in the first experiment, comprising 8 
    Gaussian clouds with 200 points each and different degrees of overlap 
    between adjacent groups.}
    \label{fig:gaussian_clouds}
\end{figure}

The first experiment evaluates the capacity of MASK to provide good 
approximations for similarity search, using a synthetic dataset 
with 8 Gaussian clouds in $\mathbb{R}^2$. This is a common benchmark 
in many previous studies on data indexing, since it is well 
characterized from a theoretical point of view \citep{Jain1988clustering,Duda2001}. 
It is possible to adjust the mean and standard deviation of each Gaussian 
distribution to control the percentage of overlap between adjacent clouds. 
As a result, we create 3 different versions of this dataset:

\begin{itemize}
    \item In the first version, the Gaussian clouds have clear separation 
    between each other, as depicted in Figure \ref{figure_gaussianCloud_noOverlap}. 
    This is a baseline test, where most data points should be correctly
    retrieved. This dataset is identified as $\mathcal{X}_{GNO}$ in experiments.
    \smallskip
    
    \item The second version represents a compromise benchmark, where
    there is a moderate overlap between contiguous Gaussian clouds, 
    as shown in Figure \ref{figure_gaussianCloud_moderateOverlap}. This
    version will be used in experiments evaluating the effect of
    the problem size or configuration parameters on MASK performance. 
    This dataset is identified as $\mathcal{X}_{GMO}$ in our evaluation tests.
    \smallskip
    
    \item The third version, depicted in Figure \ref{figure_gaussianCloud_noticeableOverlap}, consists of Gaussian clouds 
    with substantial overlap among each other. In this case, the error rate
    in point search should raise noticeably, since it is more difficult for
    MASK to retrieve data points inside overlapping regions accurately. In
    experiments, this dataset is referred to as $\mathcal{X}_{GRO}$ .
\end{itemize}

It is expected that MASK should retrieve without problems
most data points in $\mathcal{X}_{GNO}$, with clearly separated clouds.
However, as the degree of overlap between contiguous clouds increases in 
datasets $\mathcal{X}_{GMO}$ and $\mathcal{X}_{GRO}$, the error rate is expected
to grow, as well. In theory, for an exact indexing algorithm the error rate should 
tend to the Bayes (irreducible) error of the whole dataset~\citep{Duda2001}.
In this case, the classification error is determined by the 8 overlapping 
regions between contiguous clouds \citep{McLachlan1988}. However, in practice 
MASK will incur in certain additional error for exhaustive point search, on 
top of this threshold.
\smallskip

The performance of MASK in a real-world scenario is assessed in the second
experiment, using a high-dimensional and high-sparsity problem. Many competing 
algorithms for similarity search in metric spaces struggle to deal with 
datasets of this kind. The Reuters-21578 dataset~\citep{Lewis1997}, from the 
UCI repository, represents a perfect case study. This is a widely used benchmark
in supervised text classification, with a collection of more than 10,000 news 
documents, published in 1987 and categorized into 90 different topics. Some 
of these topics are very similar and present substantial overlap among them. 
Besides the results for topic categorization, this experiment measures the 
performance of MASK for indexing purposes. Henceforth, this dataset is 
identified as $\mathcal{X}_{REUT}$.

\subsection{Synthetic dataset results}
\label{subsec:experim-synthetic}

The first experiment starts comparing MASK performance against the two
opposite versions of the synthetic dataset: $\mathcal{X}_{GNO}$ (no overlap, see
Figure \ref{figure_gaussianCloud_noOverlap}) and $\mathcal{X}_{GRO}$ 
(remarkable overlap, see Figure \ref{figure_gaussianCloud_noticeableOverlap}). 
In this execution, 8 Gaussian clouds in $\mathbb{R}^2$ are generated for each dataset, 
with 200 points per cloud.  The distance function to measure dissimilarity 
between data elements is the Euclidean metric. The initial configuration parameters
for MASK are $lengthGroup = 16$ and $nCentroids = 8$, which renders a data 
summarization ratio between adjacent index levels of 2:1. In both cases, MASK 
divides the initial dataset of 1,600 points into 100 groups, with 16 points per
group. During this process, all data points are randomly assigned to one of these 
groups, so that each partition (group) may contain points from any of the 8 clouds. 
This simulates the situation in a distributed dataset, where data partitions 
are processed separately.

Once the data groups are created, we proceed to build the multilevel index. 
At the bottom level, the algorithm obtains 8 centroids per group, calculating 
a total of 800 centroids in this first layer. At the next level, the algorithm 
repeats this procedure, summarizing the 800 centroids obtained at the first level. 
Hence, in each iteration the centroids generated at the previous level are 
assigned to a new centroid at the next level above, according to the 
$k$-means algorithm and the Euclidean distance function.
\smallskip

Using this method, the algorithm eventually generates a multilevel index of 
$k$-means centroids, with a final depth of 8 layers, since the stop condition
is that the number of centroids at the top layer must be lower than $lengthGroup$.
Now, the index is ready to accept point queries that systematically search for
all points in the original dataset, following Algorithm \ref{algo:nn-search}
to seek a perfect match. In the case of dataset $\mathcal{X}_{GNO}$,
the indexing method is able to correctly find all points without 
errors, as the original clouds are well-separated between each other. 
Due to this, there is no need to perform any iterations to relocate data 
points, since the error rate cannot be further reduced.

Nevertheless, working with dataset $\mathcal{X}_{GRO}$ the error rate increases
noticeably. In this case, the multilevel index cannot retrieve elements in 
the proximity or inside overlapping regions accurately, producing an error 
rate of 40.5\% for a point search of all elements in $\mathcal{X}_{GRO}$.
\smallskip

As mentioned above, an option to reduce this error rate is relocating
data points not correctly found in the corresponding partition, reached at the 
end of the point search algorithm. After this relocation, the multilevel index is 
rebuilt and the exhaustive point search through all elements in the dataset
is repeated to obtain the new error rate. Following this approach, the 
error rate decreases from the initial 40.5\% at iteration 0 to 30.31\% 
at iteration 3. Table \ref{tab:table1} shows the error rates corresponding to each 
iteration. From these results, we can conclude that the algorithm converges 
to its best result very fast , even if it is still unable to determine the 
correct partition for some points pertaining to overlapping areas.

\begin{table}[htb]
    \caption{Change in error rate after each iteration of point relocation and index reconstruction for Gaussian clouds with remarkable overlap.\label{tab:table1}}
    \begin{center}
    \renewcommand{\arraystretch}{1.2}
        \begin{tabular}{ p{1.4cm} 
        >{\raggedleft\arraybackslash}p{0.87cm}  >{\raggedleft\arraybackslash}p{0.87cm} 
        >{\raggedleft\arraybackslash}p{0.87cm}  >{\raggedleft\arraybackslash}p{0.87cm} 
        >{\raggedleft\arraybackslash}p{0.87cm} >{\raggedleft\arraybackslash}p{0.87cm} 
        >{\raggedleft\arraybackslash}p{0.87cm} >{\raggedleft\arraybackslash}p{0.87cm}
        >{\raggedleft\arraybackslash}p{0.87cm} }
            \toprule
             Iteration & \#0 & \#1 & \#2 & \#3 &  \#4 & \#5 &  \#6 &  \#7 &  \#8 \\ \midrule
            Error rate(\%) & 40.5 & 39.06 & 32.56 & \textbf{30.31} & 32.56 & 30.62 & 34.19 & 36.25 & 36.56 \\  \bottomrule
        \end{tabular}
    \end{center}
\end{table}

\subsubsection{Influence of the dataset size}
\label{subsec:size-problem}

Since MASK is suitable for distributed computing platforms, it is crucial 
to evaluate how the problem size affects its performance. Nonetheless, 
dataset $\mathcal{X}_{GRO}$ leads to a significant error rate,
caused by the large overlap between Gaussian clouds. This large error may 
conceal the effects of other sources of variation on the index performance, 
such as the size of the dataset or changes in configuration parameters. 
For this reason, from this point we compare the performance between datasets
$\mathcal{X}_{GNO}$ (see Figure \ref{figure_gaussianCloud_noOverlap}) and $\mathcal{X}_{GMO}$ (see Figure \ref{figure_gaussianCloud_moderateOverlap}).
Different number of points per cloud are considered for both datasets
(200, 1,000, 10,000 and 100,000), in order to analyze how this variation 
affects the error rate and computation time of MASK. All simulations have 
been performed with $lengthGroup = 16$ and $nCentroids = 8$
\smallskip

\begin{figure}[ht!]
     \centering
     \subfloat[Dataset $\mathcal{X}_{GNO}$ (no overlap).\label{figure_npcN}]
         {\includegraphics[width=0.5\textwidth]{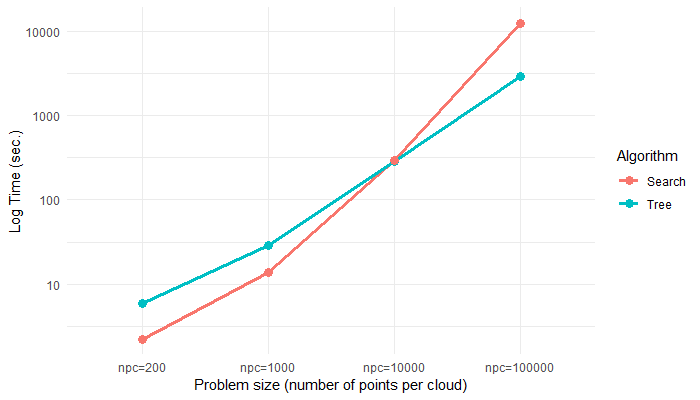}}
     \subfloat[Dataset $\mathcal{X}_{GMO}$ (moderate overlap).\label{figure_npcO}]
         {\includegraphics[width=0.5\textwidth]{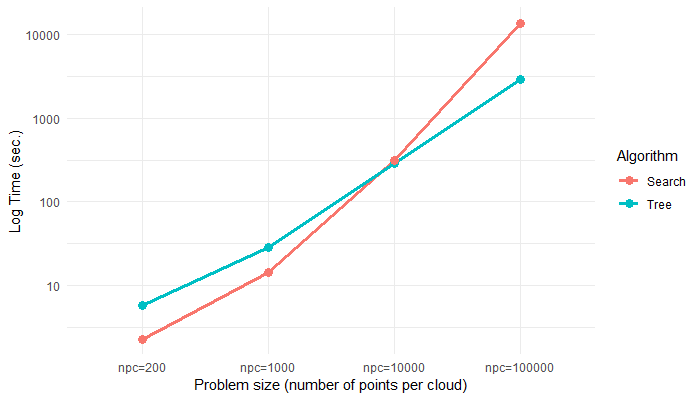}}
    \caption{Comparison of tree time and search time (log scale) for
    different values of problem size.}
    \label{fig:npc_comparison}
\end{figure}

In the case of dataset $\mathcal{X}_{GNO}$, Figure \ref{figure_npcN} displays 
the total time consumed during the index construction stage (\textit{tree time}) 
and the time required to search the whole set of points (\textit{search time}), 
for each problem size. Numerical results are presented in Table \ref{tab:table_npcN},
along with the corresponding error rates. We can conclude that the error rate 
remains very low for any size of the dataset. However, as the problem size grows,
the computation time for index construction and exhaustive point search also
increases.

\begin{table}[htb]
    \caption{Index construction time (tree time), exhaustive point 
        search duration (search time) for all elements in $\mathcal{X}_{GNO}$
        (no overlap) and error rate, with different dataset sizes. The 
        number of points per cloud in each problem (\textit{npc}) is also 
        indicated.\label{tab:table_npcN}}
    \begin{center}
    \begin{small}
    \renewcommand{\arraystretch}{1.2}
        \begin{tabular}{ >{\raggedright\arraybackslash}p{1.8cm}  
        >{\raggedleft\arraybackslash}p{2.3cm} 
        >{\raggedleft\arraybackslash}p{2.3cm} 
        >{\raggedleft\arraybackslash}p{2.3cm}
        >{\raggedleft\arraybackslash}p{2.7cm} }
            \toprule
             & Problem size $npc = 200$ & Problem size $npc = 1,000$ & Problem size $npc = 10,000$ & Problem size $npc = 100,000$ \\ 
             \midrule
             Tree time (sec.) &5.905 & 28.601 & 287.674 & 2863.450 \\ 
             
             Search time (sec.) &2.250 & 13.892 & 293.547 & 12415.074 \\
             
             Error rate(\%) &0.0 & 0.11 & 0.09 & 0.05 \\ \bottomrule
        \end{tabular}
        \end{small}
    \end{center}
\end{table}

The same analysis has been performed with dataset $\mathcal{X}_{GMO}$ (moderate
overlap). It is clear, from computation times represented in Figure \ref{figure_npcO} 
and shown in Table \ref{tab:table_npcO}, that when the problem size grows the 
time spent by MASK building the tree and searching for all points also increases. 
The error rate only experiments a small variation, that can be mainly attributed 
to the overlap between clouds. As the number of points per cloud raises, the 
overlapping region is larger, incrementing the error rate.

\begin{table}[htb]
    \caption{Index construction time (tree time), exhaustive point 
        search duration (search time) for all elements in $\mathcal{X}_{GMO}$
        (moderate overlap) and error rate, with different dataset sizes. The 
        number of points per cloud in each problem (\textit{npc}) is also 
        indicated.\label{tab:table_npcO}}
    \begin{center}
    \begin{small}
    \renewcommand{\arraystretch}{1.2}
        \begin{tabular}{ >{\raggedright\arraybackslash}p{1.8cm}  
        >{\raggedleft\arraybackslash}p{2.3cm} 
        >{\raggedleft\arraybackslash}p{2.3cm} 
        >{\raggedleft\arraybackslash}p{2.3cm}
        >{\raggedleft\arraybackslash}p{2.7cm} }
            \toprule
             & Problem size $npc = 200$ & Problem size $npc = 1,000$ & Problem size $npc = 10,000$ & Problem size $npc = 100,000$ \\
             \midrule
             Tree time (sec.) &5.842 & 28.861 & 288.439 & 2874.815 \\ 
             
             Search time (sec.) &2.284 & 14.382 & 313.715 & 13730.897 \\
             
             Error rate(\%) &11.937 & 14.212 & 14.087 & 16.334 \\ \bottomrule
        \end{tabular}
        \end{small}
    \end{center}
\end{table}

\begin{figure}[ht!]
    \centering
    \includegraphics[width=0.8\textwidth]{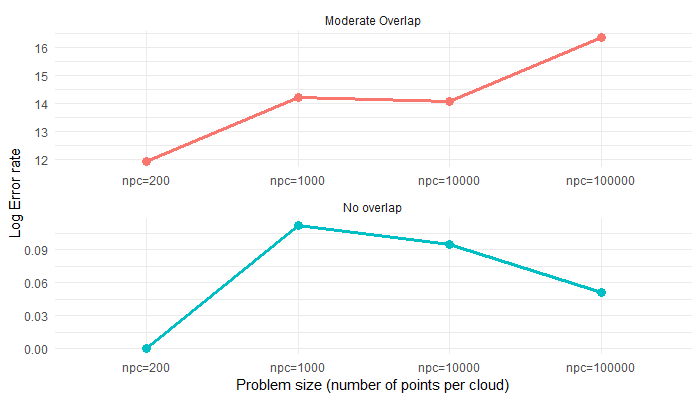}
    \caption{Comparison of the error rate when there is no overlap between dataset clouds (bottom panel) and when there is some overlap between them (top panel).}
    \label{figure_erroRate_npc}
\end{figure}

Figure \ref{figure_erroRate_npc} compares the error rate obtained with datasets
$\mathcal{X}_{GNO}$ and $\mathcal{X}_{GMO}$. As expected, the error rate is 
higher for the moderate overlap case, but it does not raise significantly for larger problem sizes with any of these two datasets. Therefore, we can conclude that the 
degree of overlap between adjacent clouds has a greater impact on the error rate 
than the size of the problem.

\subsubsection{Impact of configuration paramenters}
\label{subsec:fragmentation-tree}

The evaluation continues analyzing the influence of configuration parameters,
$lengthGroup$ and $nCentroids$, on MASK performance. Again, datasets 
$\mathcal{X}_{GNO}$ and $\mathcal{X}_{GMO}$ are used in these tests, 
fixing the number of points per cloud to 10,000 in both cases.

\begin{figure}[ht!]
    \centering
    \includegraphics[width=0.8\textwidth]{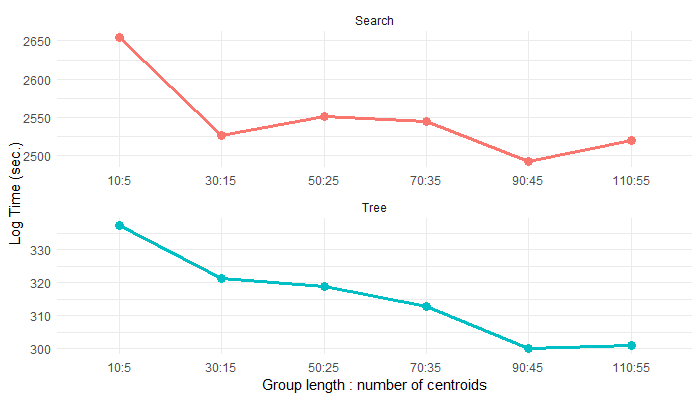}
    \caption{Computation time spent in multilevel index construction (tree) and exhaustive point query (search), for different values of $lengthGroup$ and $nCentroids$, with dataset $\mathcal{X}_{GNO}$ (10,000 points per cloud).}
    \label{figure_tgncN}
\end{figure}

Figure \ref{figure_tgncN} compares the computation time for the tree building 
phase and the exhaustive point search for dataset $\mathcal{X}_{GNO}$, using different
values of $lengthGroup$ and $nCentroids$. In all cases, the data summarization ratio
between these two configuration parameters is fixed at 2:1. As $lengthGroup$ and
$nCentroids$ increase their values, the time to build the multilevel index decreases
moderately, suggesting an inverse relationship between these magnitudes. 
At the same time, since the number of indexing levels decreases when the group 
length and the number of centroids grows, a slight reduction in search time can
be attained in some cases. Table \ref{tab:table_tgncN} explores these differences 
in more detail.

\begin{table}[ht!]
    \caption{Computation time required for multilevel index construction (\textit{tree time}) and exhaustive point query (\textit{search time}), for different values of $lengthGroup$ and $nCentroids$, with dataset $\mathcal{X}_{GNO}$ (10,000 points per cloud). The number of levels created by the index structure in each case (\textit{tree depth}) is also indicated.\label{tab:table_tgncN}}
    \begin{center}
    \renewcommand{\arraystretch}{1.2}
        \begin{tabular}{ >{\raggedright\arraybackslash}p{3.8cm}  
        >{\raggedleft\arraybackslash}p{1.8cm} 
        >{\raggedleft\arraybackslash}p{1.8cm}  
        >{\raggedleft\arraybackslash}p{2.2cm} 
        >{\raggedleft\arraybackslash}p{1.6cm} }
            \toprule
             & Tree time (sec.) & Tree depth & Search time (sec.) & Error rate(\%) \\ 
             \midrule
             \raggedright $lengthGroup=10$ $nCentroids=5$ & 337.450 & 13 & 2653.897 & 0.158 \\ 
             
             \raggedright $lengthGroup=30$ $nCentroids=15$ & 321.383 & 12 & 2525.883 & 0.125 \\
             
             \raggedright $lengthGroup=50$ $nCentroids=25$ & 318.820 & 11 & 2550.783 & 0.06 \\
             
             \raggedright $lengthGroup=70$ $nCentroids=35$ & 312.977 & 11 & 2543.880 & 0.053 \\
             
             \raggedright $lengthGroup=90$ $nCentroids=45$ & 300.070 & 10 & 2491.706 & 0.061 \\
             
             \raggedright $lengthGroup=110$ $nCentroids=55$ & 301.053 & 10 & 2518.820 & 0.055 \\
             \bottomrule
        \end{tabular}
        \end{center}
\end{table}

\begin{figure}[ht!]
    \centering
    \includegraphics[width=0.8\textwidth]{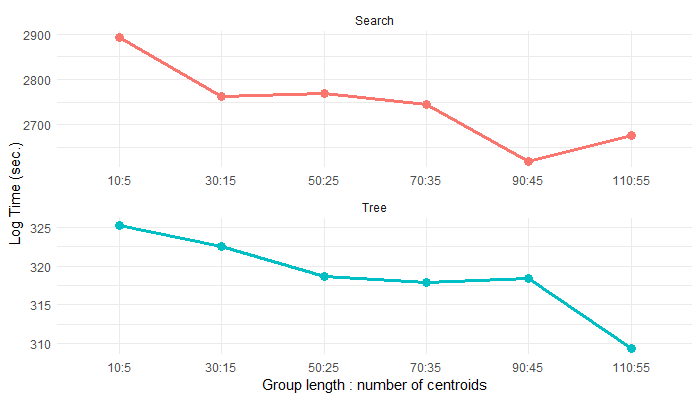}
    \caption{Computation time spent in multilevel index construction (tree) and exhaustive point query (search), for different values of $lengthGroup$ and $nCentroids$, with dataset $\mathcal{X}_{GMO}$ (10,000 points per cloud).}
    \label{figure_tgncO}
\end{figure}

A similar situation occurs when MASK is applied to dataset $\mathcal{X}_{GMO}$,
using the same series of increasing values for $lengthGroup$ and $nCentroids$ and
keeping the data summarization ratio between them at 2:1, for all cases.
Results are presented in Figure \ref{figure_tgncO} and Table \ref{tab:table_tgncO}. 
As $lengthGroup$ and $nCentroids$ are set to higher values, the computation 
time required by MASK for tree construction tends to decrease, as well as the time 
needed for exhaustive point search of the entire dataset. The values in
column \textit{tree depth} show that the number of layers in the multilevel index 
drops from 13 to 10 levels. This is the main cause behind time reduction in both
tree construction and point search. Since every individual point query must traverse 
less layers to return a result, the total time spent in exhaustive point search
is reduced.

\begin{table}[ht!]
    \caption{Numeric results for the computation time required for multilevel index construction (\textit{tree time}) and exhaustive point query (\textit{search time}), for different values of $lengthGroup$ and $nCentroids$, with dataset $\mathcal{X}_{GMO}$ (10,000 points per cloud). The number of levels created by the index structure in each case (\textit{tree depth}) is also indicated.\label{tab:table_tgncO}}
    \begin{center}
    \renewcommand{\arraystretch}{1.2}
        \begin{tabular}{ >{\raggedright\arraybackslash}p{3.8cm}  
        >{\raggedleft\arraybackslash}p{1.8cm} 
        >{\raggedleft\arraybackslash}p{1.8cm}  
        >{\raggedleft\arraybackslash}p{2.2cm} 
        >{\raggedleft\arraybackslash}p{1.6cm} }
            \toprule
             & Tree time (sec.) & Tree depth & Search time (sec.) & Error rate(\%) \\ 
             \midrule
              $lengthGroup=10$ $nCentroids=5$ & 325.294 & 13 & 2891.359 & 13.791 \\ 
             
              $lengthGroup=30$ $nCentroids=15$ & 322.588 & 12 & 2762.296 & 13.631 \\
             
              $lengthGroup=50$ $nCentroids=25$ & 318.593 & 11 & 2769.591 & 12.775 \\
             
              $lengthGroup=70$ $nCentroids=35$ & 317.898 & 11 & 2745.472 & 11.853 \\
             
              $lengthGroup=90$ $nCentroids=45$ & 318.403 & 10 & 2618.803 & 11.712 \\
             
              $lengthGroup=110$ $nCentroids=55$ & 309.307 & 10 & 2676.827 & 11.448 \\
             \bottomrule
        \end{tabular}
    \end{center}
\end{table}

\begin{figure}[ht!]
    \centering
    \includegraphics[width=0.8\textwidth]{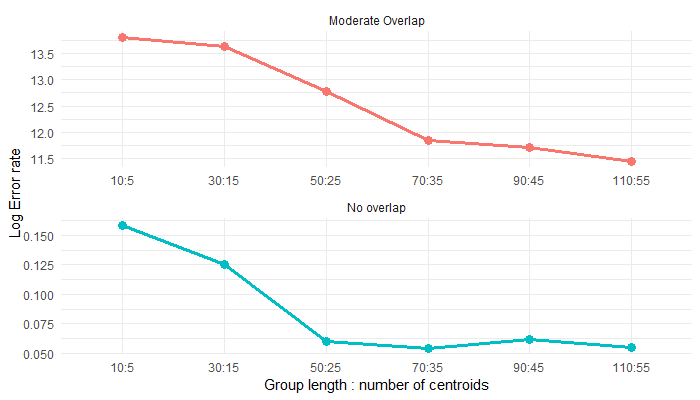}
    \caption{Comparison of changes in the error rate values with dataset $\mathcal{X}_{GNO}$ (no overlap) and $\mathcal{X}_{GMO}$ (moderate overlap),
    as $lengthGroup$ and $nCentroids$ are increased.}
    \label{figure_erroRate_tgnc}
\end{figure}

The main effect when the values of $lengthGroup$ and $nCentroids$ are increased 
is the progressive improvement of the error rate in exhaustive point search. 
Figure \ref{figure_erroRate_tgnc} compares the error rate variation for datasets
$\mathcal{X}_{GNO}$ (no overlap) and $\mathcal{X}_{GMO}$ (moderate overlap), 
respectively. Even though the error rate values for
dataset $\mathcal{X}_{GNO}$ are much lower than for $\mathcal{X}_{GMO}$, we
observe similar trends in both graphs, as $lengthGroup$ and $nCentroids$ 
values raise. From these results, it can be inferred that the influence of
both $lengthGroup$ and $nCentroids$ on the performance of MASK is consistent,
disregarding the degree of overlap between the Gaussian clouds. Index structures
with fewer layers need less time to be constructed and lead to a more precise point 
search, with lower error rate. Therefore, there is a performance advantage in
MASK when larger size of data groups (partitions) are configured. In spite of
this, after a certain point the error rate levels off, and no further improvement 
can be achieved.

\subsubsection{Role of the data summarization ratio}
\label{subsec:ratio-input-parameters}

So far, the data summarization ratio between $lengthGroup$ and $nCentroids$ 
has been kept at a fixed value of 2:1 in all experiments. Here, 
we explore how changes in this data summarization ratio affect MASK performance,
using again the $\mathcal{X}_{GNO}$ and $\mathcal{X}_{GMO}$ synthetic datasets. 
In both cases, we generate 10,000 points per cloud.
\smallskip

\begin{figure}[ht!]
    \centering
    \includegraphics[width=0.8\textwidth]{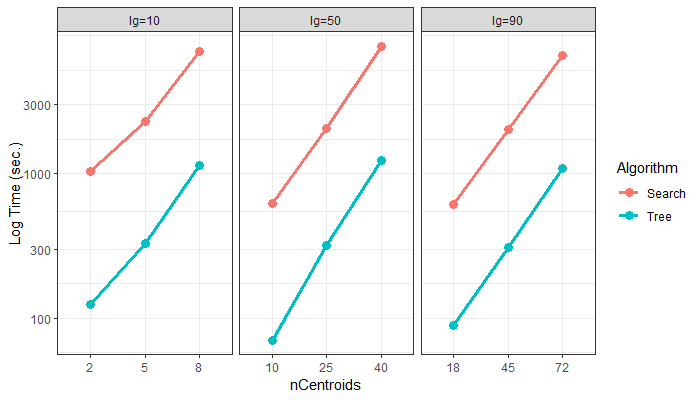}
    \caption{Computation time during multilevel index building (\textit{tree}) and exhaustive point search (\textit{red}), for different values of the data summarization ratio between $lengthGroup$ and $nCentroids$, with dataset $\mathcal{X}_{GNO}$ (10,000
    points per cloud). Each panel represents the computation time for a fixed $lengthGroup$ ($lg$) and different values of $nCentroids$.}
    \label{figure_reltgncN}
\end{figure}

Three situations are studied, corresponding to $lengthGroup = 10$, $50$ and $90$,
respectively. Figure \ref{figure_reltgncN} presents the results for $\mathcal{X}_{GNO}$
(no overlap case). The left panel display changes in computation time for tree 
assembly (blue line) and during the search process (red line), for $lengthGroup=10$. 
On the horizontal axis, $nCentroids$ gets values $2$, $5$ and $8$,
respectively. The graph shows that the computation time for both stages 
directly depends on the number of centroids per group, as expected. That is, 
if the number of centroids increases, computation time for both index construction 
and exhaustive point search also raises. The same effect is observed in the other two cases, for $lengthGroup=50$ and $90$. When $lengthGroup=50$, $nCentroids$ is set to
$10$, $25$ and $40$, and when $lengthGroup=90$ the number of centroids are $18$, 
$45$ and $72$. In this way, the same sequence of data summarization
ratios as for the case of $lengthGroup=10$ is used . The difference between computation 
time during the index construction stage and for the search process remains 
approximately constant over the three cases of $lengthGroup$.
\smallskip

\begin{figure}[ht!]
    \centering
    \includegraphics[width=0.8\textwidth]{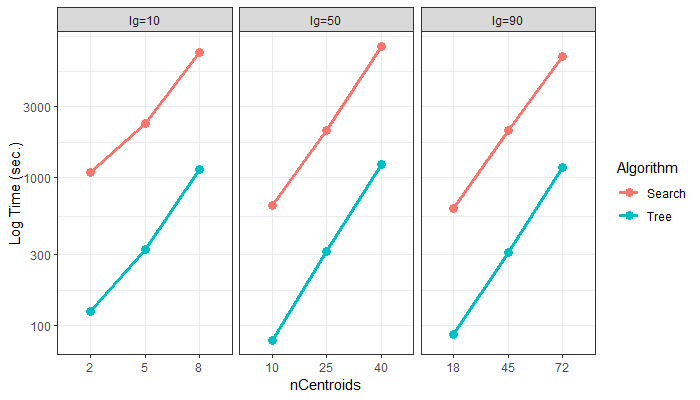}
    \caption{Computation time during multilevel index building (\textit{tree}) and exhaustive point search (\textit{red}), for different values of the data summarization ratio between $lengthGroup$ and $nCentroids$, with dataset $\mathcal{X}_{GMO}$ (10,000
    points per cloud). Each panel presents the computation time for a fixed $lengthGroup$ ($lg$) and different values of $nCentroids$.}
    \label{figure_reltgncO}
\end{figure}

Figure \ref{figure_reltgncO} displays the computation time for the moderate
overlap case (dataset $\mathcal{X}_{GMO}$). The same three scenarios 
with $lengthGroup = 10, 50$ and $90$ are
considered, with identical variation in the number of centroids for each case. 
Results are very similar to the previous ones in Figure \ref{figure_reltgncN}. 
Therefore, it can be concluded that the overlap between adjacent clouds 
does not increase computation time for tree building or search queries, when
different data summarization ratios are configured.

\begin{figure}[ht!]
    \centering
    \includegraphics[width=0.8\textwidth]{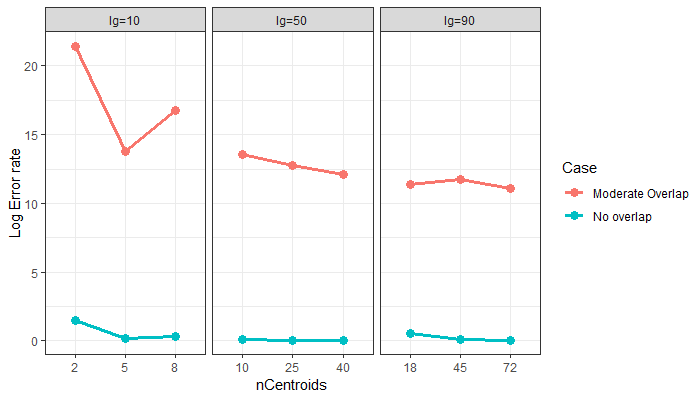}
    \caption{Comparison of error rates for datasets $\mathcal{X}_{GNO}$, without 
    no overlap between adjacent clouds (blue), and $\mathcal{X}_{GMO}$, with moderate overlap between Gaussian clouds (red).}
    \label{figure_erroRate_reltgnc}
\end{figure}

Figure \ref{figure_erroRate_reltgnc} displays changes in the error rate for
datasets $\mathcal{X}_{GNO}$ (no overlap) and $\mathcal{X}_{GMO}$ (moderate overlap) , 
in the three previous situations (with $lengthGroup = 10, 50$ and $90$), respectively. 
As seen above, the error rate decreases when $lengthGroup$ and $nCentroids$
increase their values. This difference becomes more remarkable in the moderate 
overlap case. In spite of this, if we keep the value of $lengthGroup$ constant, 
the error rate drops or remains stable when the number or centroids per group 
is increased. A satisfactory compromise solution between computation time (for
index construction and search) and error rate is found for a data summarization
ratio of 2:1.

\subsection{High-dimensional and sparse dataset results}
\label{subsec:experim-ir}

One of the most problematic aspects of MAM is that, in many cases,
they render very poor performance in high-dimensional and sparse
dissimilarity spaces. As we explained in Section \ref{intro}, this
issue is linked to the intrinsic dimensionality of our dataset, $\delta$, 
which determines the number of effective dimensions that
represent all elements. The higher the number of dimensions and the
sparsity of our dataset, the more difficult it will be for our indexing
method to effectively locate query objects.

However, a key advantage of MASK is that, if we bombard data partitions with 
enough $k$-means centroids, the indexing structure can still retrieve
data elements with good accuracy, at the expense of returning approximate
results. As shown in Section \ref{subsec:k-means-index}, when a high
number of $k$-means centroids are used for the first layer, MASK
place them according to the density of the underlying data elements. In this
way, we are able to index not only high density data regions, but also
outliers and clusters of elements separated from the main concentrations
of points.

Several experiments were conducted to evaluate the performance of MASK
in this kind of problems, using dataset $\mathcal{X}_{REUT}$. 
Individual documents are tagged according to their topic, and there exists 
substantial overlap between different categories. The goal in this evaluation 
tests is to classify documents as belonging to a particular category.

Following conventional text mining procedures~\citep{eisenstein2019introduction}, 
stop words (common words that usually do not add useful information for the
analysis), numbers, punctuation marks and white spaces were removed from documents. 
After this, the text in each document is converted to lower-case and the 
corresponding term-document matrix is created. Thus, each set of documents is 
represented as an $m \times n$ matrix, where $m$ is the number of unique terms 
in the dictionary and $n$ is the number of documents in the data set. Each 
element $w_{ij}$ of the term-document matrix represents the importance or 
weight of the term $i$ in document $j$. To obtain the value of $w_{ij}$, we 
use the TF-IDF measure \citep{aizawa2003information}, calculated through 
equation (\ref{eq:tf-idf})

\begin{equation}
\label{eq:tf-idf}
w_{ij} = tf_{ij} \times \log \left(\frac{n}{df_i} \right), 
\end{equation}
\smallskip

where $tf_{ij}$ denotes the number of occurrences of the term $i$ in 
document $j$; $n$ is the total number of documents in the dataset and 
$df_i$ represents the number of documents in which term $i$ appears.
Given these initial conditions, data elements are represented in
a dissimilarity space with many dimensions, where components are the TF-IDF 
measures for each document, and high sparsity, since many terms will be 
absent from numerous documents.
\smallskip

In a first round of experiments, documents are enconded applying TF-IDF with
stemming \citep{eisenstein2019introduction}. Results are compared with a 
second round of experiments, in which the document encoding does not include 
stemming, to assess the impact of this step on MASK search accuracy.
Several pairs of document categories have been considered for the sake of
clarity to illustrate the results: 'alum' vs. 'barley'; 'ipi vs. 'iron-steel';
'carcass' vs. 'cocoa'; 'palm-oil' vs. 'pet-chem' and 'palm-oil' vs. 'barley'.
In all cases, the input parameters for our algorithm will be $lengthGroup=16$
and $nCentroids=8$ (thus, the data summarization ratio is 2:1). The algorithm 
builds the search tree with a maximum depth of 3 levels.
\medskip

\begin{figure}[ht!]
    \centering
    \includegraphics[width=0.9\textwidth]{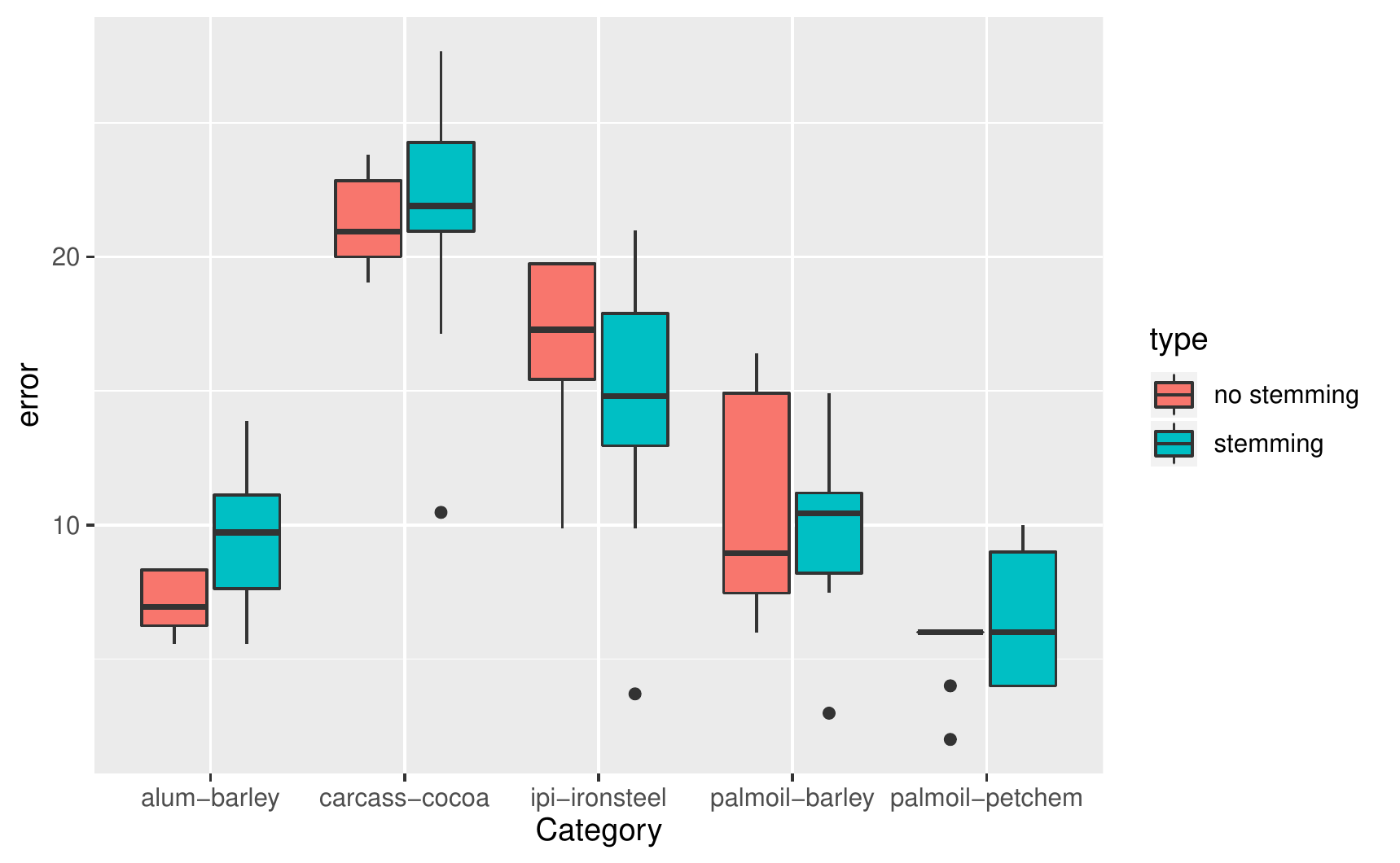}
    \caption{Boxplots comparing the classification error of documents in each pair
    of categories, obtained with and without stemming.}
    \label{figure3}
\end{figure}

Figure \ref{figure3} presents the comparison of the classification error rate 
with and without stemming, for 10 iterations of the algorithm (relocating data 
elements in each iteration, to try to improve the performance of MASK).

\begin{table}[t]
    \caption{Classification and indexing error in the best iteration of data relocation in MASK, for each pair of document categories, when stemming is applied. All simulations have been performed with $lengthGroup=16$ and $nCentroids=8$.\label{tab:table3}}
    \begin{center}
    \renewcommand{\arraystretch}{1.4}
        \begin{tabular}{l r r r r }
            \toprule
            \multirow{1}{*}{Categories} & \multicolumn{2}{c}{Classification} & \multicolumn{2}{c}{Indexing} \\ \cmidrule{2-5}
            & Best Iter. & Error & Best Iter. & Error \\ \midrule
            alum / barley & 2 & 5.55 & 1 & 8.33 \\  
            
            ipi / iron-steel & 4 & 3.7 & 0 & 12.34 \\  
            
            carcass / cocoa & 0 & 10.47 & 0 & 16.19 \\  
            
            palm-oil / pet-chem & 0 & 4.0 & 0 & 2.0 \\  
            
            palm-oil / barley & 0 & 2.98 & 1 & 17.91 \\ 
            \bottomrule
        \end{tabular}
    \end{center}
\end{table}

\begin{table}[t]
    \caption{Classification and indexing error in the best iteration of data relocation in MASK, when stemming is not applied. All simulations have been performed with $lengthGroup=16$ and $nCentroids=8$.\label{tab:table3nosteaming}}
    \begin{center}
    \renewcommand{\arraystretch}{1.4}
        \begin{tabular}{ l r r r r }
            \hline
            \multirow{1}{*}{Categories} & \multicolumn{2}{c}{Classification} & \multicolumn{2}{c}{Indexing} \\ \cmidrule{2-5}
            & Best Iter. & Error & Best Iter. & Error \\ \midrule
            alum / barley & 3 & 5.55 & 0 & 6.94 \\  
            
            ipi / iron-steel & 5 & 9.87 & 3 & 12.34 \\  
            
            carcass / cocoa & 0 & 19.04 & 0 & 15.23 \\  
            
            palm-oil / pet-chem & 0 & 2.0 & 4 & 0.0 \\  
            
            palm-oil / barley & 1 & 5.97 & 0 & 11.94 \\ 
            \bottomrule
        \end{tabular}
    \end{center}
\end{table}

Since the maximum classification error rate is less than 25\%, this confirms 
that results obtained for both cases are very accurate. However, there is a
slight improvement in several tests, when no stemming is applied. Tables
\ref{tab:table3} and \ref{tab:table3nosteaming} present the classification
and indexing error for each pair of categories when stemming is applied
and without this step, respectively. Results indicate that, in many cases, the
best iteration is either the first or the second attempt to relocate data
points, according to the previous position of centroids in the multilevel
index. Hence, in many practical situations, the actual performance gain 
attained with this iterative process is low. Due to this, it can be
assumed that the approximation provided by the initial construction of the
multilevel indexing structure is quite acceptable for high-dimensional 
problems.
\medskip

\begin{figure}[ht!]
    \centering
    \includegraphics[width=0.85\textwidth]{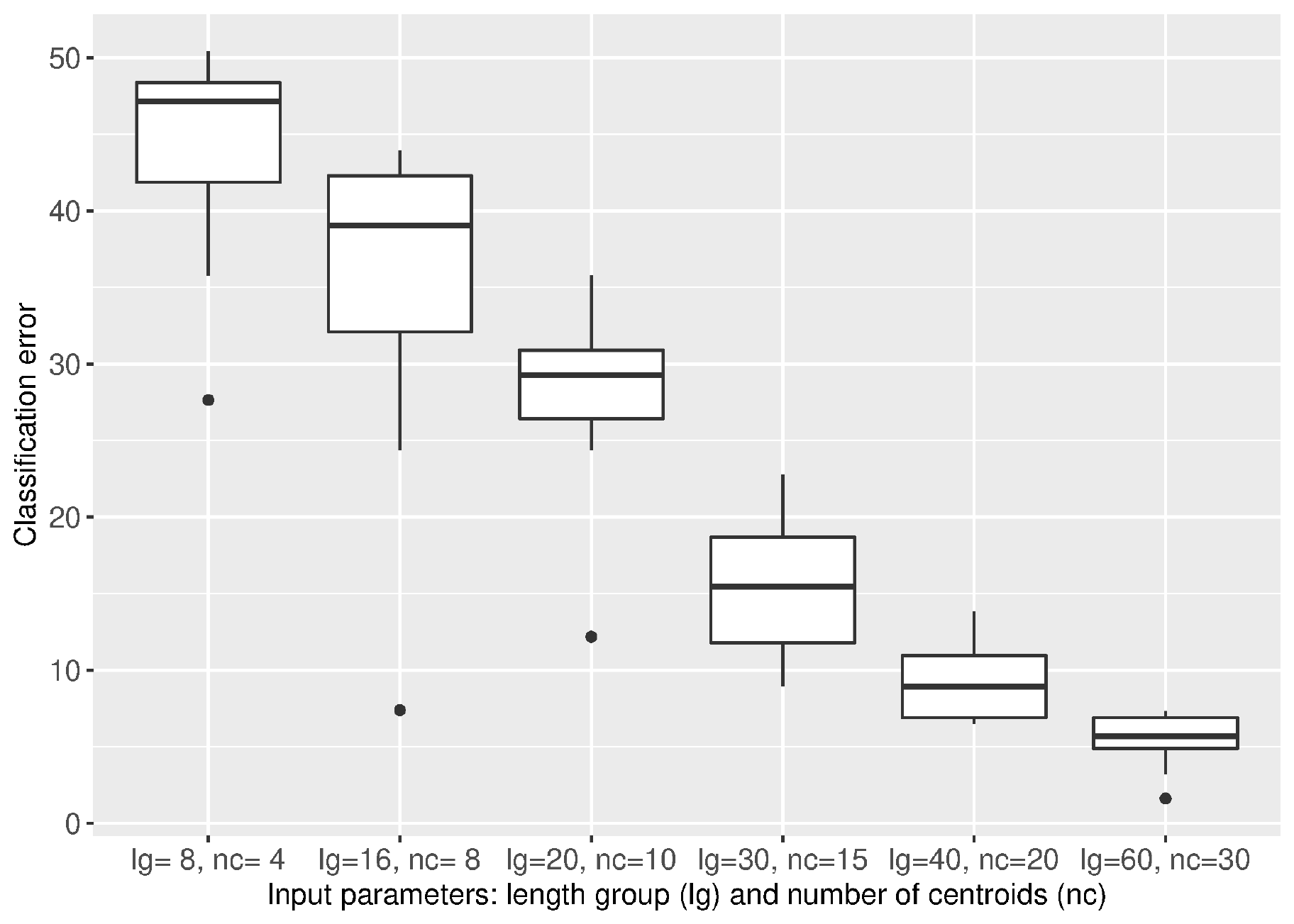}
    \caption{Classification error obtained without stemming, for categories 
    'jobs', 'iron-steel' and 'cotton' in dataset $\mathcal{X}_{REUT}$.}
    \label{figure4}
\end{figure}

Finally, we study the convergence of the algorithm as the size groups and 
the number of centroids to map each group vary. In this experiment, categories 
'jobs', 'iron-steel' and 'cotton' are considered, whereas $lengthGroup$ and 
$nCentroids$ are set to values ranging from 8-4 to 60-30, so that the 
data summarization ratio in all cases is kept at 2:1. Figure \ref{figure4} 
shows the classification error rate for each case. No stemming has been applied 
in these simulations. The classification error rate decreases when both 
input parameters increase their values. Results are quite similar when 
stemming is not applied, suggesting that this encoding step does not affect 
the accuracy of MASK, as size of data partitions and the number of centroids 
increase. Again, this is consistent with the effect of using a high number 
of centroids to map the underlying dataset, as explained in Section
\ref{subsec:k-means-index}. As a higher number of centroids is used to map 
the dataset, their location will become more advantageous for indexing purposes. 

\section{Discussion}
\label{sec:discussion}

MASK is an approximate method for similarity search based on $k$-means, 
that builds multilevel index structures suitable for different scenarios, 
from low-dimensional problems to high-dimensional and high-sparsity 
datasets in metric spaces. Algorithms to solve the principal 
types of similarity search requests can be deployed using MASK. 
Likewise, this method offers a good performance without the need to relocate 
data points after the initial index construction. The multilevel 
structure created by MASK in the first iteration already provides a 
good approximation to solve similarity queries with reasonable efficiency.

One of the main contributions of MASK is an unconventional application
of the $k$-means algorithm for information indexing, demonstrated in Section
\ref{sec:multilevel-k-means}. The classical approach in clustering problems
dictates that the number of centroids to group data points must be chosen 
beforehand. Unlike this common strategy, our approach simply suggest
that, for information indexing, one should use as many centroids 
as the infrastructure can afford. This is specially important in the lowest
layer of the multilevel index, next to the actual data points. When the
number of centroids per data partition is incremented, there is a rapid
improvement in the capacity of $k$-means to distribute centroids in
optimal locations, including denser data regions, scattered sections or
even outliers. The theoretical approach of using $k$-means for
information indexing was already considered since the inception of this
method \citep{MacQueen1967}. In addition, several algorithms for 
both exact and approximate similarity search use $k$-means to build 
their index \citep{Fukunaga1975,Ferhatosmanoglu2001,Muja2014} .
However, to the best of our knowledge, none of them follow
this unconventional application of $k$-means to map data in a flexible
manner, according to available system resources.
\smallskip

On top of this, MASK is also adequate for distributed datasets 
with partitions stored in different nodes, thanks to its bottom-up design. 
As a result, a multilevel index can be created in parallel for data partitions
in different nodes, in contrast with current approaches for similarity 
search in distributed systems ~\citep{Cech2020,Qiao2020}. In general, 
other alternative methods involve substantial data interchange 
between nodes to build the index. In MASK, index metadata 
corresponding to the top level of all nodes can be aggregated in the 
management node to select the nodes that should be involved in resolving a 
query. Thanks to its multilayer design, MASK input configuration parameters
can be tuned to achieve the desired reduction in the total number of 
centroids stored at higher levels. Indeed, as the number of layers increase, 
the set of calculated prototypes for each level is reduced by a factor 
equal to the data summarization ratio ($groupLength/nCentroids$) for that level.
\smallskip

Another advantage of MASK is that it can handle
geographically distributed datasets. For instance, consider the case
of a large organization with subsidiaries in different locations that
are geographically scattered. Using other indexing algorithms, it would
be very difficult to coalesce index metadata stored in each
location in a centralized management system. However, MASK
can also tackle this situation, aggregating the top-level centroids
from each venue. Even more complex hierarchies of data centers,
involving intermediate headquarters, can also be supported. In this case,
each intermediate center stores an aggregation
of the centroids coming from all venues under its direct oversight, and it
forward this aggregated set of centroids upstream as required.

One of the most important limitations of this new indexing method is
that it can only provide approximate results. The only warranty that MASK
can offer is that accuracy will be higher as the number of centroids
in each layer increases. Again, the data summarization ratio steers
this trade-off. The trivial scenario where one centroid is assigned
each data point (1:1 ratio) does not provide any improvement in storage
space for indexing purposes, although it renders perfect accuracy (exact 
search). Conversely, when the number of centroids per layer is gradually 
reduced, the storage gain improves, at the expense of greater loss in 
search accuracy. Specific applications should test different values for 
configuration parameters and the number of layers in the indexing structure,
to find the best fit for particular problems.

Another limitation is the need to use a powerful computing infrastructure
to store both the dataset and the index metadata described in this
approach. Nevertheless, continuous features improvements in computing
hardware and cloud architectures counteract this shortcoming to a
certain extent. Nowadays, nodes shipping large RAM units, fast
secondary storage and distributed file systems are becoming prevalent
in many organizations. This paves the way for the design of new
indexing algorithms that take advantage of extant improvements
in computing infrastructure.

In spite of these shortcomings, results from experiments show that
this new method achieves good accuracy, even in complex scenarios
involving high-dimensional and high-sparsity datasets, provided
that the problem is set in a metric space. The flexibility and 
simplicity of configuration parameters allows MASK to cover a wide 
spectrum of possible applications in many different domains.


\section{Conclusion}
\label{sec:conclusion}

In this paper we have presented MASK, a new method for approximate similarity
search based on an unconventional application of the $k$-means algorithm,
to create a multilevel index in metric spaces. 
Departing from the typical utilization of $k$-means in clustering 
problems, where the number of centroids must be specified in advance, we 
demonstrate that, for data indexing, one should use as many centroids as it 
can be afforded by the computing infrastructure, to effectively map the 
target dataset. Following this novel approach, the $k$-means algorithm 
will place the centroids at each level in convenient positions to map dense 
data regions, scattered subsets of elements or even outliers.

Moreover, the bottom-up design of this new method makes it suitable for
distributed computing systems, since each node can create
a local multilayered index in parallel, without the need of any 
data transfer between nodes. Even geographically scattered data centers
that must be managed coordinately can also benefit from using this method.
The index can be configured to meet different design goals, like storage 
capacity consumption, speed and accuracy of query resolution. This
balance between computational requirements and search performance makes
it a good candidate for a wide range of applications demanding approximate
similarity search.
\smallskip

Regarding future research on this topic, applying MASK to solve distributed
indexing problems in wireless sensor networks and other similar types of federated
and ubiquitous computing systems is a line that deserves to be explored. Recent
research in this regard ~\citep{Wan2019kmeans} emphasize the relevance of effective
indexing in this kind of technologies, that play an essential role in smart
cities, Industry 4.0 and many other settings. MASK provide clear advantages,
such as eliminating the need of data exchange among device clusters, which
reduces energy consumption. Likewise, MASK enables independent construction of 
local indexes covering specific regions of the network, that can work either 
in isolation or cooperating with each other to broaden the total coverage of
the multilevel index. This is another important asset, as the algorithm is
flexible enough to prepare an index structure that could operate in any of 
these two different modes, seamlessly changing between them.

Further experiments evaluating the performance of MASK on different datasets
must be conducted, as well. This includes assessing the performance of the
new algorithm for specific applications such as multimedia and spatial datasets, 
and  high-dimensional problems. In the last case, it will be critical to study 
the influence of the number of dimensions in the feature space over MASK indexing 
performance. In the same way, this study must be complemented with a detailed
analysis of MASK behaviour and performance using alternatives to the Euclidean
distance~\citep{Deza2013} in metric spaces. Although, in classical clustering
applications $k$-means is widely known to be tightly connected with the Euclidean
distance, in data indexing and using a large number of centroids it is possible
to adopt alternative distance functions for categorical data, strings, etc. Once
the distance function is calculated for the necessary cases, $k$-means can
build the multilevel structure as described here, based on this information.
Results from these experiments will lead to discern which combination of
distance function and initial configuration parameters is more adequate to resolve
a particular similarity search problem.

\bibliographystyle{IEEEtran}
\bibliography{MASK-App-Similarity-Search}  






\end{document}